\documentclass[pra,twocolumn,floatfix]{revtex4-1}

\usepackage{latexsym} 
\usepackage{amssymb}  
\usepackage{amsfonts} 
\usepackage{amsbsy}   
\usepackage{amsmath} 
\usepackage{graphicx}       
\usepackage{subfigure}  
\usepackage{multirow}  
\usepackage{color} 
\usepackage{bm}   
\usepackage{CJK}

\renewcommand{\>}{\rangle} 
\newcommand{\txt}{\textstyle}

\newcommand\eqn[1]{(\ref{#1})}      
\newcommand\Eqn[1]{Eq.~(\ref{#1})}  
\newcommand{\beq}{\begin{equation}}
\newcommand{\eeq}{\end{equation}}
\newcommand{\bea}{\begin{eqnarray}}
\newcommand{\eea}{\end{eqnarray}}
\newcommand{\ba}{\begin{array}}
	\newcommand{\ea}{\end{array}}

\newcommand{\half} {{\txt \frac{1}{2}}}

\newcommand{\quarter}{{\txt\frac{1}{4}}}

\newcommand{\e}{{\rm e}}   
\newcommand{\ri}{{\rm i}}
\newcommand{\rd}{{\rm d}}

\newcommand{\nn}{\nonumber}

\newcommand{\cP}{\ensuremath{\mathcal{P}}}
\newcommand{\cT}{\ensuremath{\mathcal{T}}}

\renewcommand{\Re}{{\rm Re}\,}

\newcommand{\ie}{{i.e.}}

\newcommand{\eiot}{\ensuremath{\e^{\ri \omega t}}}

\makeatletter 

\begin{document}
\title{\bf Piecewise Adiabatic Following: {General Analysis and Exactly Solvable Models}}
		
		\author{Jiangbin Gong}
		\affiliation{Department of Physics, National University of Singapore, 117542, Singapore}
		
		\author{Qing-hai Wang}
		\affiliation{Department of Physics, National University of Singapore, 117542, Singapore}

\date{\today}
	
\begin{abstract}
The dynamics of a periodically driven system whose time evolution is governed by the Schr\"{o}dinger equation with non-Hermitian Hamiltonians can be perfectly stable.  This finding was only obtained very recently and will be enhanced by many exact solutions discovered in this work. The main concern of this study is to investigate the adiabatic following dynamics in such non-Hermitian systems stabilized by periodic driving.  We focus on the peculiar behavior of stable cyclic (Floquet) states in the slow-driving limit. It is found that the stable cyclic states can either behave as intuitively expected by following instantaneous eigenstates, or exhibit piecewise adiabatic following by sudden-switching between instantaneous eigenstates.  We aim to cover broad categories of non-Hermitian systems under a variety of different driving scenarios. We systematically analyze the sudden-switch behavior by a universal route. That is, the sign change of the critical exponent in our asymptotic analysis of the solutions is always found to be the underlying mechanism to determine if the adiabatic following dynamics is trivial or piecewise.  This work thus considerably extends our early study on the same topic [Gong and Wang, Phys.~Rev.~A \textbf{97}, 052126 (2018)] and shall motivate more interests in non-Hermitian systems.

\end{abstract}
\begin{titlepage}
	\maketitle
	\renewcommand{\thepage}{}          
\end{titlepage}

\section{Introduction}

{Recent theoretical and experimental studies have touched upon many physically interesting cases described by non-Hermitian Hamiltonians \cite{BB98,benderreview,exp1,exp2}.} This class of dynamical systems is simply termed non-Hermitian systems below.   The dynamics of non-Hermitian systems can be a fruitful and useful topic considering its potential applications in understanding and controlling wave propagation with gain and loss.  From the fundamental point of view, the time evolution governed by non-Hermitian Hamiltonians  constitutes an interesting extension \cite{gongwang2010,gongwang2013} of the conventional quantum mechanics and prompts us to have a new look at a variety of quantum dynamical phenomena.

Our focus is on the general time-evolution features of non-Hermitian systems in the presence of periodic time modulation.  According to the Floquet theorem,  all important dynamical aspects of such systems with a time-periodic (though non-Hermitian) Hamiltonian can be captured by the so-called Floquet states or cyclic states.  These states are simply the eigenstates of the one-period time evolution operator and hence simply acquire an overall phase factor after one period, with the phase, called Floquet eigenphase, being complex in general.  Interestingly, it was earlier shown that by tuning the system parameters, some peculiar dynamical regime can emerge, where periodic time modulation may help to stabilize non-Hermitian systems because all cyclic states can be made to possess real eigenphases \cite{J2014,gongwang2015}.  Indeed, under such circumstances,  the stroboscopic time evolution becomes unitary up to a similarity transformation \cite{gongwang2015} and hence resembles much to the familiar quantum evolution governed by Hermitian Hamiltonians.

To gain more insights into non-Hermitian systems stabilized by periodic time modulation, we aim to carry out a rather systematic and  technical study of these systems in the limit of slow time modulation.  In particular,  can such systems just comfortably adapt to the slowly-time-varying Hamiltonian, just as what is na\"ively suggested by the adiabatic theorem from the conventional quantum mechanics \cite{adiabatic-theory} and classical mechanics?
This issue is also of general interest because in Nature, slow and almost periodic modulation is often naturally introduced to a broad class of systems around us,  by the slow periodic change of the four seasons.  The common wisdom would say Yes to the question here, but caution must be taken because even in the domain of conventional quantum mechanics, new understandings of the physics of adiabatic following are still emerging \cite{adiabatic-theory,pertubation-theory,zhangqiprl,zhangqinjp,longwenprb,longwenprl}. Indeed, as shown by a recent study by us \cite{gongwang2018}, the concept of adiabatic following with the instantaneous eigenstates of the system Hamiltonian is actually not necessarily true. Instead, a non-Hermitian system can display unexpected behavior of sudden switching between different instantaneous eigenstates of the system Hamiltonian. That is, contrary to our na\"ive expectations, the adiabatic following dynamics is piecewise.  For several two-level non-Hermitian systems subject to a harmonic driving,  the system's time-evolution was analyzed from a geometrical point of view, via the projected Hilbert space depicted by the Bloch sphere \cite{gongwang2018}. During the time evolution, the system's trajectory on the Bloch sphere may display drastic changes, a phenomenon unique in non-Hermitian systems. {We stress that this sudden switch behavior is not caused by circling around a spectral exceptional point (EP) because all the dynamics are chosen to be away from and not circling around any EP. It is also a distinctly different phenomenon from the unstable evolution when the initial state is chosen to be an energy eigenstate \cite{BU,Rotter,UCF,Hailong}.} To demonstrate this, we actually went a long way by adopting a physical and highly useful geometrical phase concept \cite{Berry84,AA87,Berryphase-PRL} to characterize the exotic dynamical behavior \cite{gongwang2018}.
This geometrical approach has recently led to a new scheme to characterize the so-called dynamical phase transitions in non-Hermitian systems \cite{longwenarxiv}.

The task of this work is to lay a solid theoretical foundation, as well as a framework, for understanding the above-mentioned sudden-switch behavior in the adiabatic following dynamics of non-Hermitian systems. To that end, we  treat a broad class of non-Hermitian two-level systems periodically modulated by one, two or even three harmonic driving fields.  We shall analyze in detail when and how the sudden-switch phenomenon occurs, thus demonstrating that the intriguing sudden-switching behavior in the adiabatic following dynamics can be typical.  How different driving schemes impact on the adiabatic following dynamics will become clear from this work.  More than one sudden-switches (sometimes as many as four) within one driving period are also found to be possible.   Our careful theoretical calculations can identify clearly the underlying critical boundaries in the parameter space of such systems.   Indeed, having and not having sudden-switch behavior in the adiabatic following dynamics represent two distinctively different dynamical features. {Because each of the two cases corresponds to certain regions in the parameter space,  each region can be viewed as a phase (analogous to the widely used terminology of spontaneous \cP\cT\ symmetry breaking in Refs.~\cite{BB98,benderreview,exp1,exp2} when referring to crossing different parameter regions).}
As seen below, even when the dynamics is not exactly solvable,  there is still a powerful technique that allows us to carry out necessary asymptotic analysis in the slow-driving limit.  The rather general treatment in these systems not only extend the models we studied before \cite{gongwang2015,gongwang2018}, but can also cover interesting models studied by others \cite{BU,UCF}.   Our comprehensive results shall become a useful reference for any future theoretical and experimental study of adiabatic following dynamics in periodically driven and non-Hermitian systems.

Section \ref{sec:theory} outlines some general treatments and our notation for $2\times2$ non-Hermitian Hamiltonians periodically modulated in time.  Based on the nature of the periodic time dependence of the Hamiltonian we divide our model systems into four categories,  three of them respectively treated in Sec. III, IV, and V.   The last section concludes this work.   Appendix A presents some details of an intermediate result and Appendix B treats a more technical case where the driving field has three different frequency components including a constant term.

\section{Generic $2\times2$ time-dependent Hamiltonians}
\label{sec:theory}
Consider the time-dependent Schr\"odinger equation
\begin{equation}
\ri\hbar|\dot\Psi(t)\> = H(t)|\Psi(t)\>,
\end{equation}
where the overhead dot denotes the time derivative, $\dot{f}(t) = \frac{\rd f(t)}{\rd t}$. The time-dependent Hamiltonian is assumed to be a $2\times2$ matrix, whose most general form is given by
\begin{equation}
H(t) = \left(
\begin{array}{cc}
f_0(t) + f_3(t) &	f_1(t) - \ri f_2(t)\\
f_1(t) + \ri f_2(t)&	f_0(t) - f_3(t)
\end{array}
\right).
\end{equation}
with all the components $f_\mu(t)$ with $\mu=0,1,2,3$ being complex-valued functions of time in general.

\subsection{Trace can always be gauged away}
Consider a gauge transformation
\begin{equation}
|\Psi(t)\> \equiv \eta(t)|\psi(t)\>,
\end{equation}
where
\begin{equation}
\eta (t) \equiv \exp\left[ \frac{1}{\ri\hbar} \int^t \rd\tau f_0(\tau)\right].
\end{equation}
The Schr\"odinger equation satisfied by $|\psi(t)\>$ is
\begin{equation}
\ri\hbar|\dot\psi(t)\> = h(t)|\psi(t)\>,
\label{eqn:Schr}
\end{equation}
where
\begin{eqnarray}
h(t) &\equiv& H(t) - \ri\hbar \eta^{-1}(t)\dot\eta(t) \openone
\nn\\
&=& H(t) - f_0(t) \openone\nn\\
&=& \left(
\begin{array}{cc}
f_3(t) 		&	f_1(t)-\ri f_2(t)\\
f_1(t) +\ri f_2(t) 	&	-f_3(t)
\end{array}
\right) \nn\\
&\equiv& \mathbf{f}(t)\cdot\bm{\sigma},
\label{eqn:h(t)}
\end{eqnarray}
with $\openone$ to be the unity matrix, $\bm{\sigma} = (\sigma_1,\sigma_2,\sigma_3)$ to be Pauli matrices, and the dot product is defined as
\begin{equation}
\mathbf{A}\cdot \mathbf{B} \equiv A_1B_1+A_2B_2 + A_3B_3.
\end{equation}
Certainly, the dot product between two complex vectors is in general complex valued as well. Both its real and imaginary parts are invariant under rotations of the vectors. Upon this representation change, the transformed Hamiltonian $h(t)$ becomes traceless. For this reason,  in the following we will only consider the traceless Hamiltonians without loss of generality.

\subsection{Instantaneous eigenstates}
The instantaneous eigenstates of the traceless Hamiltonian in \Eqn{eqn:h(t)} are found to be
\begin{equation}
|E_\pm(t)\> = \left(
\begin{array}{c}
f_1(t)-\ri f_2(t)\\
 - f_3(t) \pm \sqrt{\mathbf{f}(t)\cdot \mathbf{f}(t)}
\end{array}\right),
\label{eqn:instantaneous}
\end{equation}
with the corresponding eigenvalues
\begin{equation}
E_\pm(t) = \pm\sqrt{\mathbf{f}(t)\cdot \mathbf{f}(t)}.
\end{equation}
{Clearly, the exceptional point (EP) is located at $\mathbf{f}(t)\cdot \mathbf{f}(t)=0$.}

\subsection{Differential equations}
Let us rewrite the two-component wavefunction as follows:
\begin{equation}
|\psi(t)\> = \left(
\begin{array}{c}
a(t) \\
b(t)
\end{array}\right).
\end{equation}
Then the Schr\"{o}dinger equation yields (in $\hbar=1$ unit)
\begin{eqnarray}
\ri \dot{a}(t) &=& f_3(t) a(t) + \left[f_1(t)-\ri f_2(t) \right] b(t),
\label{eqn:1stOrderA}\\
\ri \dot{b}(t) &=&  \left[f_1(t)+\ri f_2(t) \right] a(t) - f_3(t) b(t).
\label{eqn:1stOrderB}
\end{eqnarray}
If $f_1 (t)-\ri f_2(t)$ is non-zero or only vanishes at isolated points, we may cancel $b(t)$ and obtain a second order differential equation for $a(t)$,
\begin{equation}
\ddot{a}-\frac{\dot{f}_1 - \ri \dot{f}_2}{f_1 - \ri f_2} \dot{a} + \left(\mathbf{f}\cdot\mathbf{f} - \ri \frac{\dot{f}_1 - \ri \dot{f}_2}{f_1 - \ri f_2} f_3 +\ri \dot{f}_3 \right)a =0.
\label{eqn:a(t)}
\end{equation}
Here we have suppressed the time variable $t$. Similarly, if $f_1 (t)+\ri f_2(t)$ is not vanishing over a time interval, a very similar equation of $b(t)$ can be obtained,
\begin{equation}
\ddot{b}-\frac{\dot{f}_1 + \ri \dot{f}_2}{f_1 + \ri f_2} \dot{b} + \left(\mathbf{f}\cdot\mathbf{f} + \ri \frac{\dot{f}_1 + \ri \dot{f}_2}{f_1 + \ri f_2} f_3 - \ri \dot{f}_3 \right)b =0.
\label{eqn:b(t)}
\end{equation}
From now on, we assume that $f_1 (t)-\ri f_2(t)$ vanishes at most at isolated points and solve for \Eqn{eqn:a(t)}. In the case that $H(t)$ has a lower triangular form with $f_1 (t)-\ri f_2(t)=0$, one simply needs to solve \Eqn{eqn:b(t)} first and then follow a similar procedure described below. In the very special case with both $f_1 (t)-\ri f_2(t)=0$ and $f_1 (t)+\ri f_2(t)=0$, we have $f_1(t)=f_2(t)=0$ over a time interval and then $H(t)$ reduces to a diagonal form. In that almost trivial case $a(t)$ and $b(t)$ can be easily found by solving the two first-order differential equations as suggested by Eqs.~(\ref{eqn:1stOrderA}) and (\ref{eqn:1stOrderB}).

If \Eqn{eqn:a(t)} is solvable, with the two linearly independent special solutions given by $y_1(t)$ and $y_2(t)$, then the general solution $a(t)$ has the form
\begin{equation}
a(t) = C_1 y_1(t) + C_2 y_2(t).
\end{equation}
Equation \eqn{eqn:1stOrderA} then directly gives $b(t)$:
\begin{equation}
b(t) = \alpha(t) a(t) + \beta(t) \dot{a}(t),
\label{eqn:bt}
\end{equation}
with
\begin{equation}
\alpha(t) \equiv -\frac{f_3}{f_1 - \ri f_2} \quad \mathrm{and} \quad \beta(t) \equiv \frac{\ri}{f_1 - \ri f_2}.
\label{eqn:alphabeta}
\end{equation} Needless to say, only for some very special forms of $\mathbf{f}(t)$, the explicit solutions to \Eqn{eqn:a(t)} can be indeed found. In the following sections, we will discuss some of these special cases. However, even without obtaining the explicit solution, we can still proceed with our discussions assuming their existence.

\subsection{Time evolution operator}
For a given initial state, one can match it with the general solution
$\left(\begin{array}{c}
a(t) \\
b(t)
\end{array}\right)$ to find $C_1$ and $C_2$.
That is, using
\begin{equation}
\left(\begin{array}{c}
a(0)\\
b(0)
\end{array}\right)=\left(\begin{array}{c}
a_0\\
b_0
\end{array}\right),
\end{equation}
we find the two constants of integrations to be
\begin{eqnarray}
C_1 &=& \frac{ a_0 \left[ \alpha(0) y_2(0) + \beta(0) y_2'(0)\right] - b_0 y_2(0)}{\beta(0) \mathcal{W}(0)}, \nn\\
C_2 &=& -\frac{ a_0 \left[ \alpha(0) y_1(0) + \beta(0) y_1'(0)\right] - b_0 y_1(0)}{\beta(0) \mathcal{W}(0)},
\end{eqnarray}
where $\alpha(t)$ and $\beta(t)$ are defined above in \Eqn{eqn:alphabeta} and
we have introduced the Wronskian of the two special solutions,
\begin{equation}
\mathcal{W}(t) \equiv y_1(t) y_2'(t) - y_1'(t)y_2(t).
\end{equation}

Let $U(t)$ be the time evolution operator. By definition we have
\begin{equation}
\left(\begin{array}{c}
a(t)\\
b(t)
\end{array}\right)\equiv U(t) \left(\begin{array}{c}
a(0)\\
b(0)
\end{array}\right).
\end{equation}
Using the previous expressions for $a(t)$ and $b(t)$, we arrive at
\begin{equation}
U(t) = \frac{1}{\beta(0)\mathcal{W}(0)} \left(\begin{array}{cc}
\tilde{U}_{11}(t) & \tilde{U}_{12}(t)\\
\tilde{U}_{21}(t) & \tilde{U}_{22}(t)
\end{array}\right),
\end{equation}
where
\begin{eqnarray}
\tilde{U}_{11}(t) &=&  \left[ \alpha(0) y_2(0) + \beta(0) y_2'(0) \right] y_1(t) \nn\\
&&\quad - \left[\alpha(0) y_1(0) + \beta (0) y_1'(0) \right] y_2(t), \nn\\
\tilde{U}_{12}(t) &=& -y_2(0) y_1(t) + y_1(0) y_2(t), \nn\\
\tilde{U}_{21}(t) &=&  \left[\alpha(0) y_2(0) + \beta(0) y_2'(0) \right] \left[\alpha(t) y_1(t) + \beta(t) y_1'(t) \right]  \nn\\
&&\quad -\left[\alpha(0) y_1(0) + \beta(0) y_1'(0) \right] \left[\alpha(t) y_2(t) + \beta(t) y_2'(t) \right], \nn\\
\tilde{U}_{22}(t) &=& - y_2(0) \left[\alpha(t) y_1(t) + \beta(t)  y_1'(t) \right]\nn\\
&&\quad + y_1(0) \left[\alpha(t) y_2(t) + \beta(t) y_2'(t) \right] .
\end{eqnarray}
Interestingly, if we rewrite the arbitrary initial state in terms of the following superposition,
\begin{eqnarray}
\left(\begin{array}{c}
a(0)\\
b(0)
\end{array}\right) &=& C_1 \left(\begin{array}{c}
y_1(0)\\
\alpha(0) y_1(0) + \beta(0) y_1'(0)
\end{array}\right) \nn\\
&&\quad + C_2 \left(\begin{array}{c}
y_2(0)\\
\alpha(0) y_2(0) + \beta(0) y_2'(0)
\end{array}\right),
\label{eqn:ICs}
\end{eqnarray}
then the time-evolved state at time $t$ does maintain this form, namely,
\begin{eqnarray}
\left(\begin{array}{c}
a(t)\\
b(t)
\end{array}\right)&=& U(t) \left(\begin{array}{c}
a(0)\\
b(0)
\end{array}\right)\nn\\
 &=& C_1 \left(\begin{array}{c}
y_1(t)\\
\alpha(t) y_1(t) + \beta(t) y_1'(t)
\end{array}\right)
\nn\\
&&\quad
 + C_2 \left(\begin{array}{c}
y_2(t)\\
\alpha(t) y_2(t) + \beta(t) y_2'(t)
\end{array}\right),
\label{eqn:wavefunction}
\end{eqnarray}
with all time-dependent functions updated but with precisely the same $C_1$ and $C_2$ as two integration constants.
This result can be regarded as one direct outcome of the linearity of the Schr\"{o}dinger equation, a feature useful for our analysis below.

In the slow-driving limit, if one of the two special solutions used above is much greater than the other, then the wavefunction approximately parallel to the dominant term on the right-hand side of \Eqn{eqn:wavefunction}. For example, suppose $y_1(t)\gg y_2(t)$ for some $t$, then the term proportional to $C_2$ in \Eqn{eqn:wavefunction} is negligible. During the evolution, this dominance relation may change due to the well-known Stokes phenomenon. Say, for some other $t$, $y_1(t)\ll y_2(t)$, then the $C_2$ term in \Eqn{eqn:wavefunction} becomes dominant. This can then lead to a drastic change in the time-evolving state when it is projected onto smoothly changing basis states, such as the instantaneous eigenstate representation.  This qualitative understanding will be important  when we analyze different models in the following sections \cite{footnote1}.

\begin{figure}[h!]
	\begin{center}
		\includegraphics[width=0.45\columnwidth]{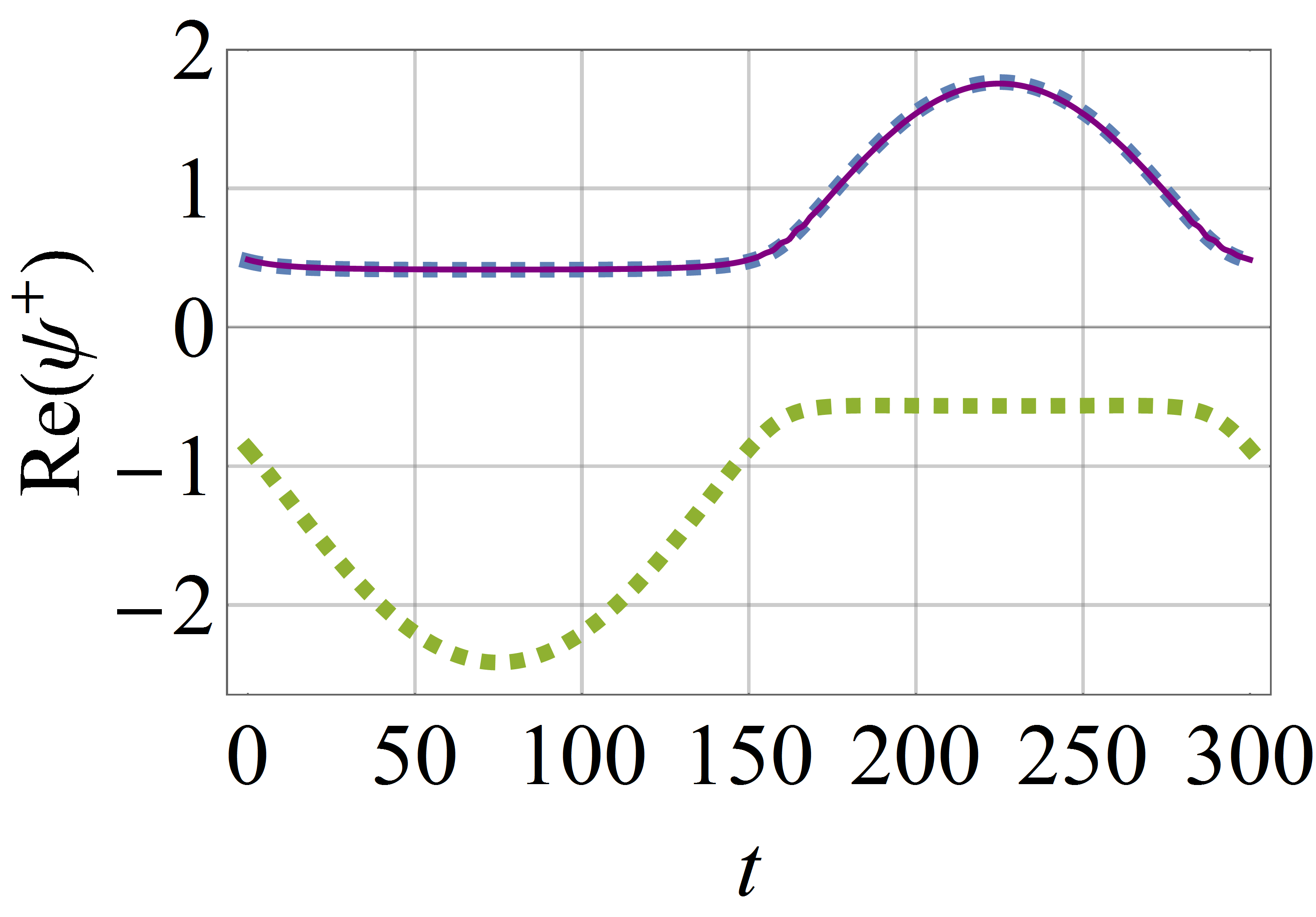}
		\includegraphics[width=0.45\columnwidth]{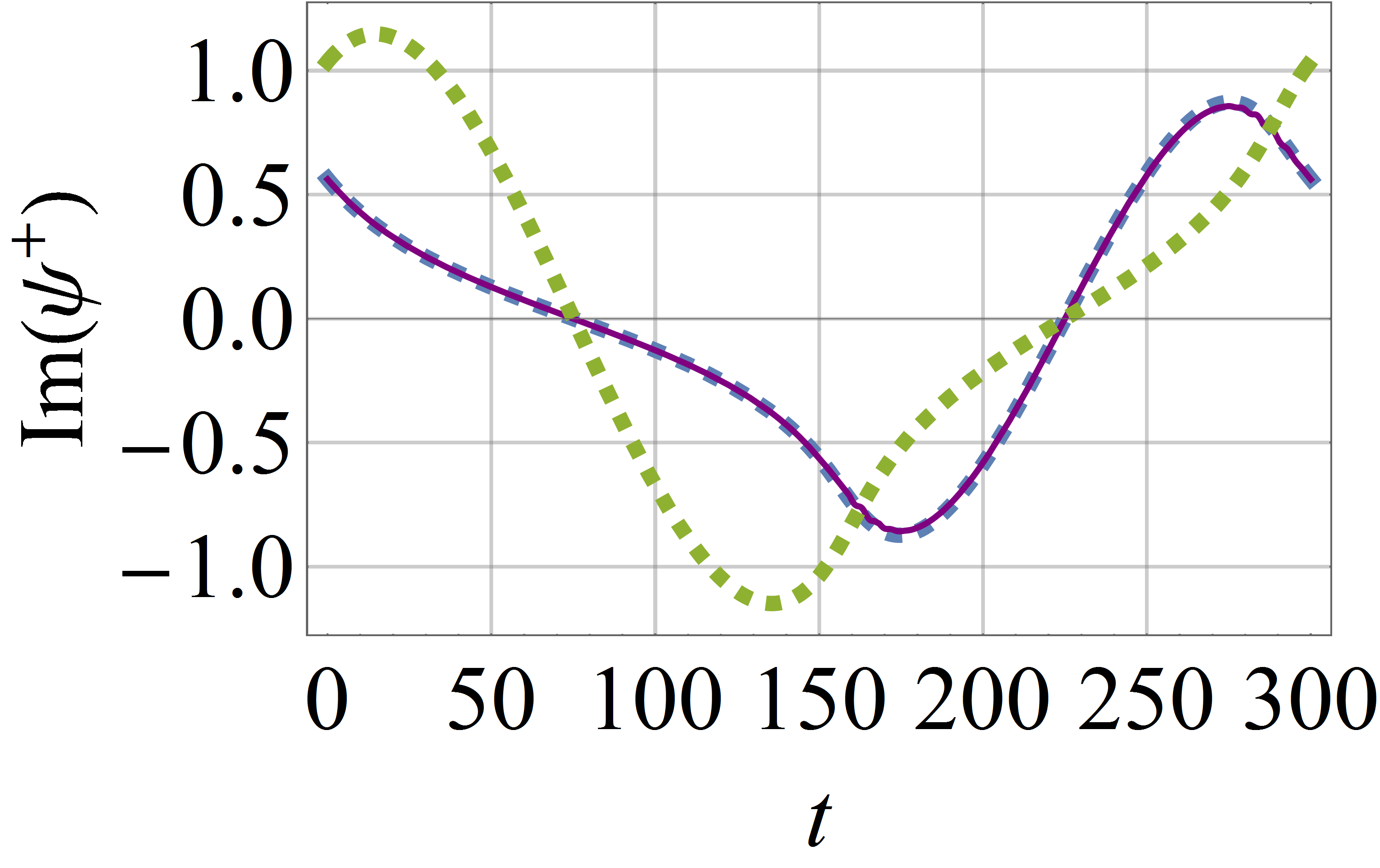}
		\includegraphics[width=0.45\columnwidth]{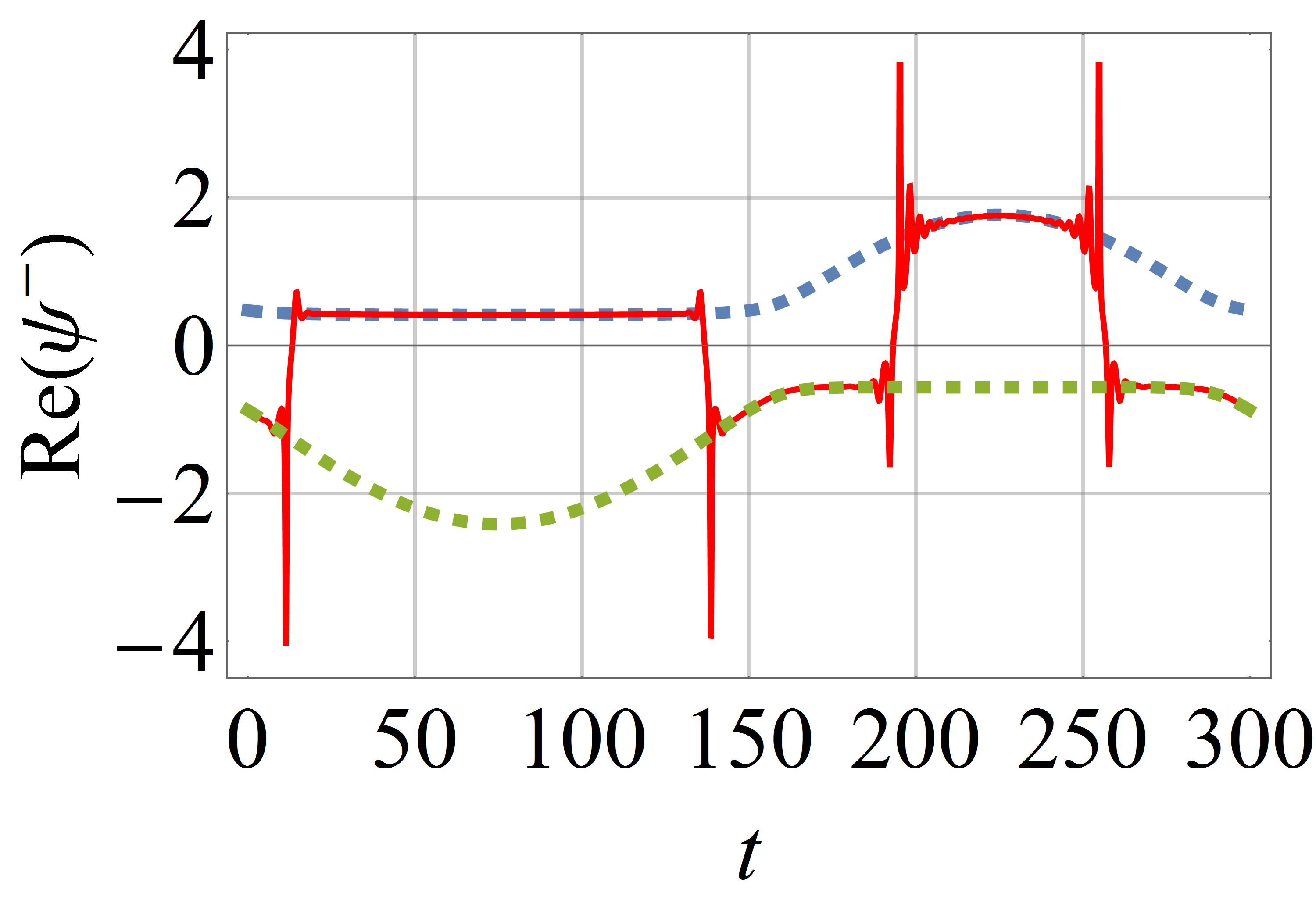}
		\includegraphics[width=0.45\columnwidth]{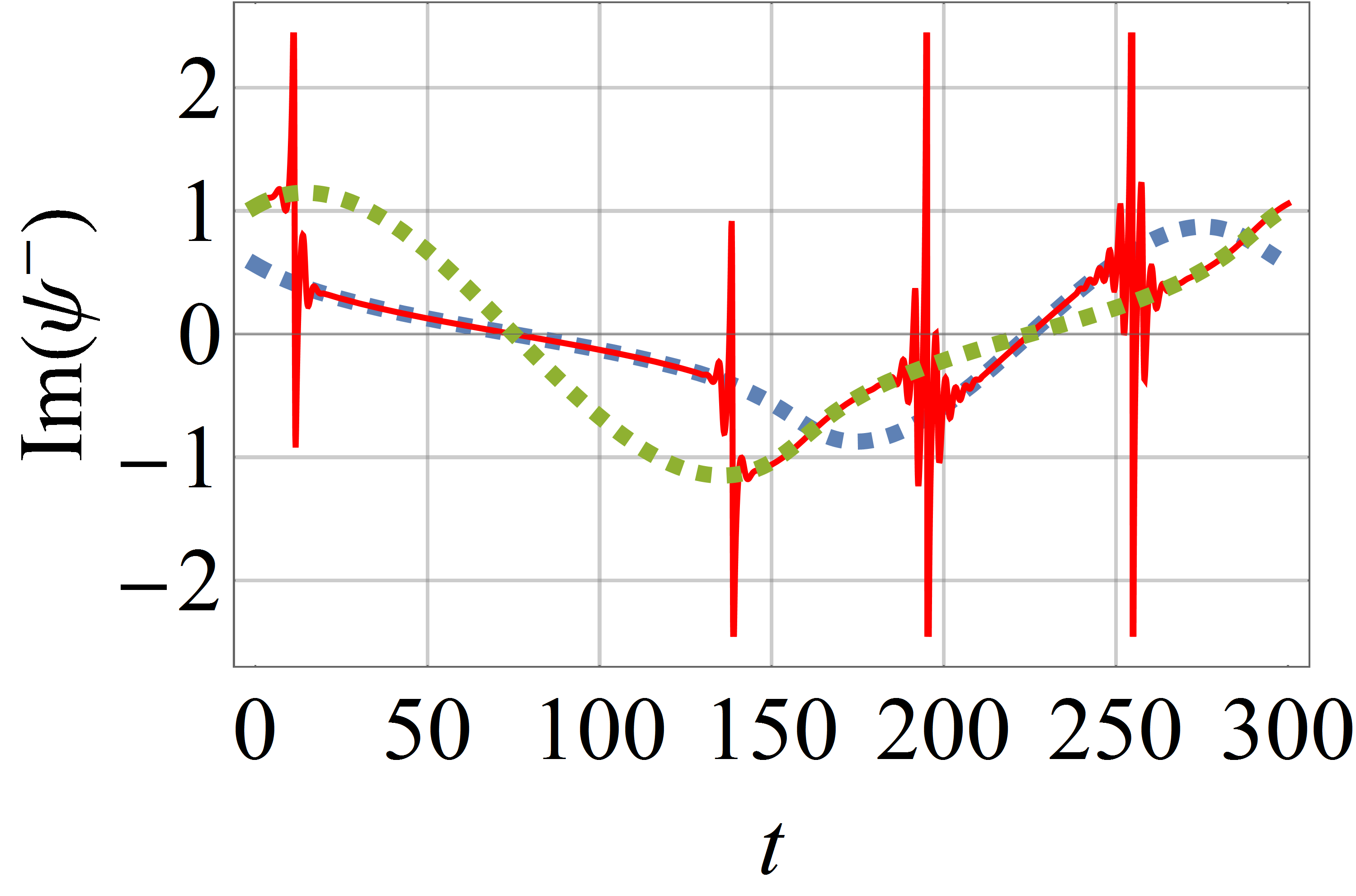}
		\caption{\label{fig:UCF}(color online)
		Real and imaginary parts of $\psi(t)=b(t)/a(t)$ for a time-evolving states (solid lines) and the instantaneous energy eigenstates (dashed lines) for the UCF model in \Eqn{eqn:HUCF} with $r=0.2\ri$, $\rho=-0.8$, and $T=300$.  The top two panels show that one Floquet state, $|F^+(t)\>$ adiabatically follows one of the instantaneous energy eigenstates. The bottom panels present the piecewise following by the other Floquet state, $|F^-(t)\>$. It hops four times within one period. {Here and in other figures, all plotted quantities are
		in dimensionless units with $\hbar=1$.}}
	\end{center}
\end{figure}

Figure \ref{fig:UCF} shows two distinct behaviors for time-evolving states under the slow-driving limit. The top two panels show that one state follows one of the instantaneous energy eigenstates. The bottom two panels present a drastically different feature. In most of the time, the time-evolving state follows the instantaneous energy eigenstates in a piecewise fashion. During some relative short time windows, the time-evolving state ``hops'' from one instantaneous eigenstate to the other. {Note that the two energy eigenstates have not swapped, which evidently means that no EP has been circled.} This profound breakdown of the adiabatic theorem in non-Hermitian systems was first reported by us \cite{gongwang2018}.

\subsection{Time-Periodic systems}
Consider now time-periodic Hamiltonians with,
\begin{equation}
H(t+T)=H(t).
\end{equation}
According to the Floquet theorem, the eigenstates of the time evolution operator $U(T)$ of one period are cyclic states (also called Floquet states), with
\begin{equation}
U(T) |F^\pm(0)\> = \e^{\pm \ri \phi} |F^\pm(0)\>.
\label{eqn:Floquet}
\end{equation}
The zero sum of the two phases is a result of the traceless feature of the Hamiltonian we can always assume. As such, if $\phi\neq 0 $ and $\phi\neq \pm \pi $, then we get two non-degenerate cyclic states whose time-dependence is determined by $ |F^\pm(t)\> =U(t) |F^\pm(0)\>$. Note that $\phi$ is complex valued in general.  When $\phi$ is real and nonzero, the system has long-term stability insofar the system can only acquire a phase factor on the unit circle after an arbitrary number of driving periods. In this sense, the system possesses ``extended unitarity'' according to Ref.~\cite{gongwang2015}.  The special situation with degenerated eigenphases must be treated carefully. If $U(T)$ has two distinct eigenstates, the system is still stable. If $U(T)$ is not diagonalizable, only one cyclic state may be obtained. To form a complete set, one must find a generalized eigenstate in the Jordan chain. A generic state is a linear combination of the Floquet state and the generalized eigenstate, which leads to a linear growth in time \cite{Longhi}. To avoid such a complication, we only consider non-degenerated Floquet states in the rest of the paper.

Note also that the Floquet states can always be expanded in the manner depicted by \Eqn{eqn:ICs} with time-independent coefficients $C_1$ and $C_2$.

\section{Models with a single Fourier component}
All components $f_\mu(t)$ parameterizing a periodic Hamiltonian $H(t)$ are complex-valued functions with the same period,
\begin{equation}
T\equiv\frac{2\pi}{\omega}.
\end{equation}
From the Fourier theorem, we can write $\mathbf{f} (t)$ in terms of its Fourier components, i.e.,
\begin{equation}
\mathbf{f} (t) = \sum_{n=-\infty}^{\infty} \tilde{\mathbf{f}}_n \e^{\ri n \omega t}.
\end{equation}
For completeness, we start from the simplest cases with just one nonzero Fourier component.

\subsection{Time-independent Hamiltonian}
When the Hamiltonian is time-independent, say
\begin{equation}
\mathbf{f}(t) = \mathbf{p},
\end{equation}
we have
\begin{equation}
H_0 = \left(
\begin{array}{cc}
p_3 		&	p_1-\ri p_2\\
p_1 +\ri p_2 	&	-p_3
\end{array}
\right).
\end{equation}
If $p_1-\ri p_2\neq 0$, the solution is simply
\begin{eqnarray}
a(t) &=& C_1 \e^{\ri \sqrt{\mathbf{p}\cdot\mathbf{p}} t} + C_2 \e^{-\ri \sqrt{\mathbf{p}\cdot\mathbf{p}} t}, \nn\\
b(t) &=& -C_1 \frac{p_3 + \sqrt{\mathbf{p}\cdot\mathbf{p}}}{p_1 -\ri p_2} \e^{\ri \sqrt{\mathbf{p}\cdot\mathbf{p}} t} - C_2 \frac{p_3 - \sqrt{\mathbf{p}\cdot\mathbf{p}}}{p_1 -\ri p_2} \e^{-\ri \sqrt{\mathbf{p}\cdot\mathbf{p}} t}.\nn\\
\end{eqnarray}
Fitting the initial conditions in \Eqn{eqn:ICs} gives
\begin{eqnarray}
C_1 &=& \frac{1}{2\sqrt{\mathbf{p}\cdot\mathbf{p}}}\left[\left(\sqrt{\mathbf{p}\cdot\mathbf{p}} - p_3\right) a(0) -\left(p_1 - \ri p_2\right)b(0)\right],\nn\\
C_2 &=& \frac{1}{2\sqrt{\mathbf{p}\cdot\mathbf{p}}}\left[\left(\sqrt{\mathbf{p}\cdot\mathbf{p}} + p_3\right) a(0) +\left(p_1 - \ri p_2\right)b(0)\right].\nn\\
\end{eqnarray}
Since the Hamiltonian is time-independent, the cyclic states coincide with the energy eigenstates in \Eqn{eqn:instantaneous},
\begin{equation}
|F^\pm(0)\> = |E^\pm\> = \left(
\begin{array}{c}
p_1-\ri p_2\\
 - p_3 \pm \sqrt{\mathbf{p}\cdot \mathbf{p}}
\end{array}\right).	
\end{equation}
The overall phases of the time-evolving state $|F^\pm(t)\> $ are simply the dynamical phases (time integration of the eigenvalue of the Hamiltonian),
\begin{equation}
|F^\pm(t)\> = \e ^ {\mp \ri \sqrt{\mathbf{p}\cdot\mathbf{p}} t} |E^\pm\>.
\end{equation}
The system is stable if the eigenenergy is real. {Note that real energy eigenvalues do not necessarily mean that the Hamiltonian is Hermitian. Rather, such kind of real eigenvalues can be due to the $\mathcal{PT}$ symmetry \cite{BB98,benderreview,BBJ,WCZ}. The Floquet states $|F^\pm(t)\>$ are always parallel with the corresponding energy eigenstates $|E^\pm\>$.  One can safely say that not much is interesting in the dynamics.

\subsection{Single-frequency driving}
When the Hamiltonian contains only one Fourier component, say
\begin{equation}
\mathbf{f}(t) = \mathbf{p}\e^{\ri n \omega t}.
\end{equation}
For any nonzero $n$, we can always rescale $\omega\to\omega/n$ to absorb the parameter $n$. Thus, without loss of generality, consider a Hamiltonian with the following single frequency driving,
\begin{equation}
H_1 = \left(
\begin{array}{cc}
p_3 		&	p_1-\ri p_2\\
p_1 +\ri p_2 	&	-p_3
\end{array}
\right)\eiot.
\end{equation}
If $p_1-\ri p_2\neq 0$, the solution is
\begin{eqnarray}
a(t) &=& C_1 \exp\left(\frac{\sqrt{\mathbf{p}\cdot\mathbf{p}}}{\omega}\eiot  \right) + C_2 \exp\left(-\frac{\sqrt{\mathbf{p}\cdot\mathbf{p}}}{\omega}\eiot  \right) , \nn\\
b(t) &=& -C_1 \frac{p_3 + \sqrt{\mathbf{p}\cdot\mathbf{p}}}{p_1 -\ri p_2} \exp\left(\frac{\sqrt{\mathbf{p}\cdot\mathbf{p}}}{\omega}\eiot  \right) \nn\\
&&\quad - C_2 \frac{p_3 - \sqrt{\mathbf{p}\cdot\mathbf{p}}}{p_1 -\ri p_2} \exp\left(-\frac{\sqrt{\mathbf{p}\cdot\mathbf{p}}}{\omega}\eiot  \right).
\end{eqnarray}
Matching this solution to the initial condition as in \Eqn{eqn:ICs}, one obtains
\begin{eqnarray}
C_1 &=& \frac{\left(p_3 + \sqrt{\mathbf{p}\cdot\mathbf{p}}\right) a(0) + \left(p_1 - \ri p_2\right)b(0)}{2\sqrt{\mathbf{p}\cdot\mathbf{p}}} \exp\left(\frac{\sqrt{\mathbf{p}\cdot\mathbf{p}}}{\omega}\right),\nn\\
C_2 &=&- \frac{\left(p_3 - \sqrt{\mathbf{p}\cdot\mathbf{p}}\right) a(0) +\left(p_1 - \ri p_2\right)b(0)}{2\sqrt{\mathbf{p}\cdot\mathbf{p}}} \exp\left(-\frac{\sqrt{\mathbf{p}\cdot\mathbf{p}}}{\omega}\right). \nn\\
\end{eqnarray}
For this time-dependent Hamiltonian subject to an overall time-dependent factor, it is straightforward to find that
the Floquet states again coincide with the instantaneous energy eigenstates, with
\begin{equation}
|F^\pm(0)\> = |E_\pm(0)\> = \left(
\begin{array}{c}
p_1-\ri p_2\\
- p_3 \pm \sqrt{\mathbf{p}\cdot\mathbf{p}}
\end{array}\right),
\end{equation}
and 
\begin{eqnarray}
|F^\pm(t)\> &=& U(t)|F^\pm(0)\> \nn\\
&=& \exp\left[\pm \frac{\sqrt{\mathbf{p}\cdot\mathbf{p}}}{\omega} (1-\eiot)\right]\left(
\begin{array}{c}
p_1-\ri p_2\\
- p_3 \pm \sqrt{\mathbf{p}\cdot\mathbf{p}}
\end{array}\right).	\nn\\
\end{eqnarray}
That is, as time evolves the state only acquires some overall complex-valued phases.
After one period, each cyclic state returns to itself with total phase zero.  The system is hence stable. Interestingly,  even though the Hamiltonian is time-dependent, the cyclic state always aligns with the eigenstates of the Hamiltonian, no matter how fast or slow the driving is. Once again, we see no rich dynamics here.

\section{Models with one Fourier component as well as a zero-frequency term}
\subsection{General considerations}
In this section, we consider models with two Fourier components with one of the two being a constant. For example,
\begin{equation}
\mathbf{f}(t) = \mathbf{p} + \mathbf{q}\,\e^{\ri n \omega t}.
\end{equation}
We may also rescale $\omega\to\omega/n$ to absorb the parameter $n$. Thus, without loss of generality, let us consider
\begin{eqnarray}
H_{01} &=& \mathbf{p}\cdot\bm{\sigma} + \mathbf{q}\cdot\bm{\sigma} \e^{\ri \omega t}\nn\\
&=& \left(
\begin{array}{cc}
p_3 		&	p_1-\ri p_2\\
p_1 +\ri p_2 	&	-p_3
\end{array}
\right) + \left(
\begin{array}{cc}
q_3 		&	q_1-\ri q_2\\
q_1 +\ri q_2 	&	- q_3
\end{array}
\right) \eiot. \nn\\
\end{eqnarray}
The two components satisfy the differential equations
\begin{eqnarray}
\ri \dot{a}(t) &=& \left(p_3 + q_3 \e^{\ri \omega t}\right) a(t) + \left[(p_1-\ri p_2) + (q_1-\ri q_2) \e^{\ri \omega t} \right] b(t),\nn\\
\ri \dot{b}(t) &=&  \left[(p_1+\ri p_2) + (q_1+\ri q_2) \e^{\ri \omega t} \right] a(t) - \left(p_3 + q_3 \e^{\ri \omega t}\right) b(t). \nn\\
\end{eqnarray}
If one of $(p_1-\ri p_2)$ and $(q_1-\ri q_2)$ does not vanish, we may cancel $b(t)$ and obtain a second order differential equation for $a(t)$,
\begin{eqnarray}
&&\left[(p_1-\ri p_2) + (q_1-\ri q_2) \e^{\ri \omega t} \right] \ddot{a} (t) -\ri\omega (q_1-\ri q_2) \e^{\ri \omega t} \dot{a}(t) \nn\\
&=&-\left\{(p_1-\ri p_2) \mathbf{p}\cdot\mathbf{p} \right.\nn\\
&& + \left[ (p_1-\ri p_2)(2\mathbf{p}\cdot\mathbf{q} - \omega q_3) + (q_1-\ri q_2)(\mathbf{p}\cdot\mathbf{p} + \omega p_3)\right]\eiot \nn\\
&& + \left[(p_1-\ri p_2) \mathbf{q}\cdot\mathbf{q} +2 (q_1-\ri q_2)  \mathbf{p}\cdot\mathbf{q}\right] \e^{2\ri \omega t} \nn\\
&& \left. + (q_1-\ri q_2)  \mathbf{q}\cdot\mathbf{q}\,\e^{3\ri \omega t}
\right\}a(t).
\end{eqnarray}
This equation can be separated into two parts, one is proportional to $(p_1-\ri p_2)$, and the other is proportional to $ (q_1-\ri q_2) \e^{\ri \omega t}$,
\begin{equation}
(p_1-\ri p_2)\{\text{eqn}_P\} +  (q_1-\ri q_2)\eiot \{\text{eqn}_Q \}=0,
\label{eqn:master01}
\end{equation}
where
\begin{eqnarray}
\text{eqn}_P &\equiv& \ddot{a} (t)  + \left[ \mathbf{p}\cdot\mathbf{p} + (2\mathbf{p}\cdot\mathbf{q} - \omega q_3) \eiot + \mathbf{q}\cdot\mathbf{q}\, \e^{2\ri \omega t} \right] a(t), \nn\\
\text{eqn}_Q &\equiv& \ddot{a} (t) -\ri\omega \dot{a}(t) + \left[ (\mathbf{p}\cdot\mathbf{p} + \omega p_3) + 2 \mathbf{p}\cdot\mathbf{q}\, \eiot \right.\nn\\
&&\quad \left. + \mathbf{q}\cdot\mathbf{q}\, \e^{2\ri \omega t} \right] a(t). \nn
\end{eqnarray}
This ``master" equation is solvable if either $p_1-\ri p_2$ or $q_1-\ri q_2$ vanishes.

\subsection{Asymptotic analysis}
\label{sec:exponent}
To proceed we now introduce a change of variable (which will be used in other following sections as well). In particular, we define
\begin{equation}
Z(t)\equiv \eiot \quad {\rm and}\quad a(Z) = a(t).
\end{equation}
Equation (\ref{eqn:master01}) then becomes
\begin{eqnarray}
&&- (p_1-\ri p_2) \left\{\omega^2  Z^2 a''(Z)  + \omega^2  Za'(Z) \right.\nonumber\\
&&\qquad \left. +\left[\mathbf{p}\cdot\mathbf{p} + (2\mathbf{p}\cdot\mathbf{q} - \omega q_3) Z +\mathbf{q}\cdot\mathbf{q}\,Z^2  \right] a(Z)  \right\}\nn\\
&=& (q_1-\ri q_2)Z \left\{ Z^2\omega^2 a''(Z) \right. \nn\\
&&\quad \left.+  \left[ \mathbf{p}\cdot\mathbf{p} + \omega p_3 + 2 \mathbf{p}\cdot\mathbf{q}\, Z + \mathbf{q}\cdot\mathbf{q}\,Z^2\right]a(Z) \right\}.
\label{eqn:masterz}
\end{eqnarray}

Of our central interest is always the slow-driving limit.  {However, one must be careful because  $\omega\to0$ is a singular limit of \Eqn{eqn:masterz}. Na\"ively setting $\omega=0$ will produce only one but not two solutions for $a(Z)$. The proper procedure is to first let \cite{BenderBook}}
\begin{equation}
a(Z) \equiv \e^{\frac{m(Z)}{\omega}}.
\end{equation}
Then $m(Z)$ satisfies
\begin{eqnarray}
&& \left[(p_1-\ri p_2) + (q_1-\ri q_2) Z \right] Z^2 \left\{\omega m''(Z) + \left[m'(Z)\right]^2 \right\} \nn\\
&&\quad  + \omega  (p_1-\ri p_2) Z m'(Z) \nn\\
&=& \left[(p_1-\ri p_2) + (q_1-\ri q_2) Z \right] \left( \mathbf{p}\cdot\mathbf{p} + 2\mathbf{p}\cdot\mathbf{q}\,Z +\mathbf{q}\cdot\mathbf{q}\,Z^2 \right)\nn\\
&&\quad -\omega \left[ (p_1-\ri p_2)q_3 - (q_1-\ri q_2) p_3 \right] Z .
\label{eqn:masterm}
\end{eqnarray}
This equation has a smooth slow-driving limit.  {Namely, there are two distinct solutions.} To the leading order,
\begin{equation}
Z^2 \left[m'(Z)\right]^2 \sim  \mathbf{p}\cdot\mathbf{p} + 2\mathbf{p}\cdot\mathbf{q}\,Z +\mathbf{q}\cdot\mathbf{q}\,Z^2 ,\qquad  \omega\to0.
\end{equation}
Thus,
\begin{equation}
m(Z)\sim \pm\int^Z\rd x\frac{\sqrt{\mathbf{p}\cdot\mathbf{p} + 2\mathbf{p}\cdot\mathbf{q}\,x +\mathbf{q}\cdot\mathbf{q}\,x^2}}{x}.
\label{eqn:m(z)}
\end{equation}
This indefinite integral has a closed form (see \textbf{2.261}, \textbf{2.266}, and \textbf{2.267} of Ref.~\cite{RG}),
\begin{eqnarray}
&&\int\rd x\frac{\sqrt{a + bx +cx^2}}{x} \nn\\
&=& \sqrt{a + bx +cx^2} - \frac{\sqrt{a}}{2} \ln\frac{2a +bx + 2\sqrt{a} \sqrt{a + bx +cx^2}}{2a +bx - 2\sqrt{a} \sqrt{a + bx +cx^2}} \nn\\
&&\quad + \frac{b}{4\sqrt{c}} \ln\frac{b+2cx + 2\sqrt{c} \sqrt{a + bx +cx^2}}{b+2cx - 2\sqrt{c} \sqrt{a + bx +cx^2}}.
\label{eqn:integral1}
\end{eqnarray}
The integration constant is chosen such that the integral vanishes when $\sqrt{a + bx +cx^2}=0$.

We are now ready to analyze what happens to $a(Z)= \e^{\frac{m(Z)}{\omega}}$ in the slow-driving limit. Since $m(Z)$ is obtained to the leading order of $\omega$, the behavior of $a(Z)$ is now determined by $m(Z)$ obtained in
\Eqn{eqn:m(z)}. For convenience in referring to the solution of $m(Z)$, we define $g(\omega t)$ to be the right hand side of Eq.~(\ref{eqn:integral1}) with
\begin{equation}
a= \mathbf{p}\cdot\mathbf{p},\quad b= 2\mathbf{p}\cdot\mathbf{q}, \quad c=\mathbf{q}\cdot\mathbf{q}
\end{equation}
and $x=\eiot$. Because $m(Z)$ defined above has $\pm$ solutions,  asymptotically, $a(t)$ must be a linear combination of the two following terms, {
\begin{equation}
a(t) \sim D_1 v(\omega t) \e^{\frac{g(\omega t)}{\omega}} + D_2 v(\omega t) \e^{-\frac{g(\omega t)}{\omega}}, \quad \omega\to 0,
\end{equation} 	
where $v(\omega t)$ is an unimportant prefactor which can be found by calculating the next order correction for $m(Z)$ in \Eqn{eqn:m(z)}.} Note that the constant pair $D_1$ and $D_2$ may not be the same as the pair $C_1$ and $C_2$ in \Eqn{eqn:ICs}. Suppose that the real part of $g(\omega t)$ flips its sign at $t=0$. For example, $\Re [g(\omega t)]>0$ when $t<0$ and $\Re [g(\omega t)]<0$ when $t>0$. As long as neither $D_1$ nor $D_2$ vanishes, then
\begin{equation}
a(t) \sim \left\{
	\begin{array}{ll}
	D_1 v(\omega t) \e^{\frac{g(\omega t)}{\omega}}, &\quad t<0,\\
	D_2 v(\omega t) \e^{-\frac{g(\omega t)}{\omega}}, &\quad t>0.
	\end{array}
\right.
\end{equation}	
Because of the large factor of $\frac{1}{\omega}$ in the exponent in the slow driving limit, $\omega\to 0$, the flip of this asymptotic behavior can occur in a \textit{relatively} short time window. This observation makes it intriguing to understand the existence and features of such a sudden-switch or hopping phenomenon.

{Figure \ref{fig:UCF_g} illustrates the real part of the critical exponent $g(\theta)$ for the same model used in Fig.~\ref{fig:UCF}. In this particular example, $\Re[g(\theta)]$ changes signs four times. As a result, one of the Floquet states hops four times during one period in the
slow-driving limit, as shown in Fig.~\ref{fig:UCF}. }

\begin{figure}[h!]
	\begin{center}
		\includegraphics[width=0.5\columnwidth]{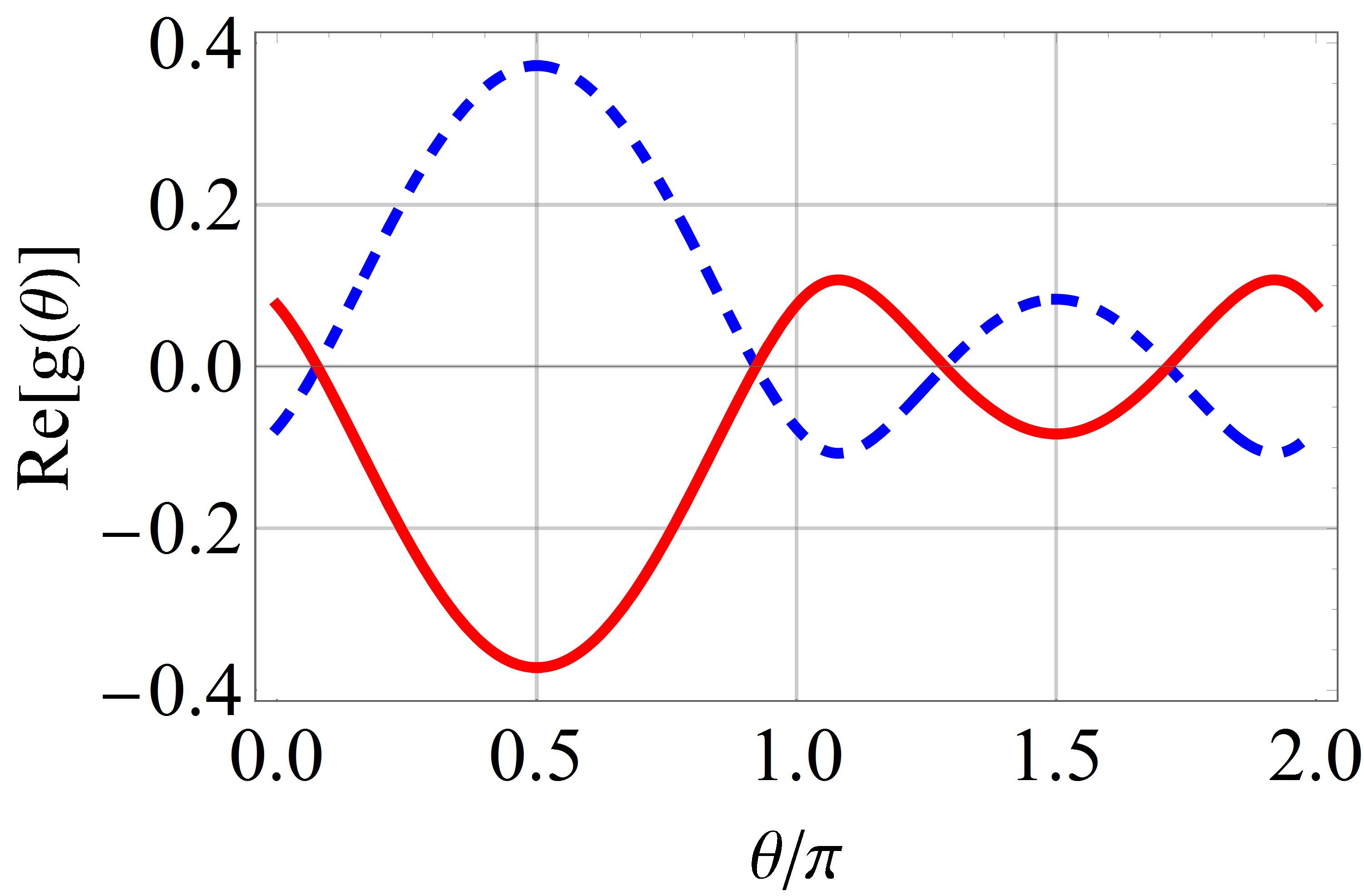}
		\includegraphics[width=0.42\columnwidth]{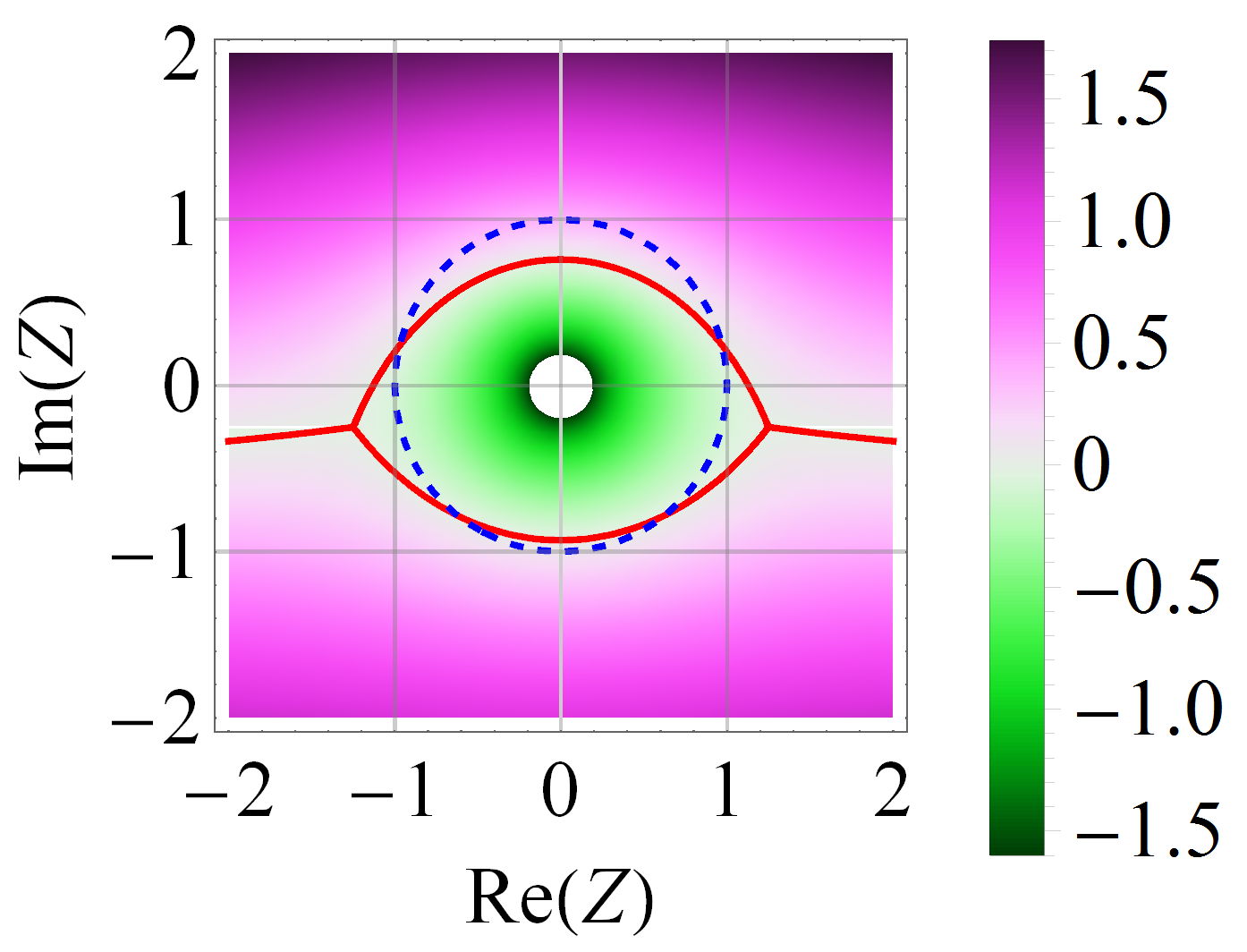}
		\caption{\label{fig:UCF_g}(color online) Real part of the critical exponent $g(\theta)$ for the model in \Eqn{eqn:HUCF} with $r=0.2\ri$ and $\rho=-0.8$. Left panel: $\pm\Re[g(\theta)]$ as functions of $\theta = \omega t$. Note that $\Re[g(\theta)]$ vanishes four times in one period, at precisely the same time when hopping occurs in Fig.~\ref{fig:UCF}. Right panel: a density plot of $\Re[g(\theta)]$ on the complex $Z$-plane. The (red) solid lines are the Stokes lines, where $\Re[g(\theta)]=0$. The (blue) dashed line is the unit circle $Z=\e^{\ri\theta}$. In one period, $\theta$ increases from $0$ to $2\pi$. Hopping occurs when the two types of lines intersect.}
	\end{center}
\end{figure}

\subsection{Exactly solvable cases after a time-independent rotation}
Equation~(\ref{eqn:masterz}) can be simplified dramatically and becomes solvable if either $(p_1-\ri p_2)$ or $(q_1-\ri q_2)$ vanishes. Remarkably, this requirement can be always fulfilled by a time-independent rotation (see Appendix \ref{sec:appendixA} for the details).

\subsection{Solvable case A with $p_1=\ri p_2$}
\label{sec:SFI}
In this case,
\begin{equation}
\mathbf{p}\cdot\mathbf{p} = p_3^2.
\end{equation}
The Hamiltonian has the form
\begin{equation}
H_{\rm 01A}(t) = \left(
\begin{array}{cc}
p_3+q_3\,\eiot & (q_1-\ri q_2)\eiot\\
2p_1+(q_1 + \ri q_2) \eiot & -p_3-q_3\,\eiot
\end{array}\right).
\label{eqn:H_I}
\end{equation}
Equation~(\ref{eqn:masterz}) then reduces to
\begin{equation}
\omega^2  Z^2a''(Z) =\left[(p_3 + \omega )p_3 + 2 \mathbf{p}\cdot\mathbf{q}\,Z +  \mathbf{q}\cdot\mathbf{q}\,Z^2\right]a(Z).
\label{eqn:Whittaker}
\end{equation}

\subsubsection{$\mathbf{q}\cdot\mathbf{q}\neq0$}
If $\mathbf{q}\cdot\mathbf{q}$ does not vanish, by changing variables
\begin{equation}
z(t) \equiv 2\frac{\sqrt{\mathbf{q}\cdot\mathbf{q}}}{\omega} \eiot \quad \mathrm{and} \quad
a(z) \equiv a(t),
\label{eqn:60}
\end{equation}
we get a Whittaker's equation (see \textbf{13.14.1} of Ref.~\cite{DLMF}),
\begin{equation}
a''(z) + \left(-\frac{1}{4} -\frac{\mathbf{p}\cdot\mathbf{q}}{\omega z \sqrt{\mathbf{q}\cdot\mathbf{q}}} -\frac{p_3(p_3+\omega)}{\omega^2 z^2}\right) a(z) =0.
\end{equation}
The solutions are
\begin{equation}
a(z) = C_1 W_{\kappa,\mu} \left(z\right) + C_1 M_{\kappa,\mu} \left(z\right)
\end{equation}
with
$
\kappa \equiv -\frac{\mathbf{p}\cdot\mathbf{q}}{\omega\sqrt{\mathbf{q}\cdot\mathbf{q}}}$, $
\mu \equiv \frac{1}{2} + \frac{p_3}{\omega}.
$
In terms of the new variable $z$, we have
\begin{eqnarray}
b(z) &=& \alpha(z) a(z) + \beta a'(z)\nn\\
&=& C_1 \left[ \alpha(z) W_{\kappa,\mu}(z) + \beta W_{\kappa,\mu}'(z) \right] \nn\\
&&\quad
+ C_2\left[ \alpha(z) M_{\kappa,\mu}(z) + \beta M_{\kappa,\mu}'(z) \right],
\end{eqnarray}
where we have introduced short-handed notations
\begin{eqnarray}
\alpha(z) &\equiv& -\frac{2p_3\sqrt{\mathbf{q}\cdot\mathbf{q}}}{\omega z(q_1-\ri q_2)} - \frac{q_3}{q_1-\ri q_2}, \nn\\
\beta &\equiv& -2\frac{\sqrt{\mathbf{q}\cdot\mathbf{q}}}{q_1-\ri q_2}. \nn
\end{eqnarray}

At $t=T$, $z=z_0\e^{2\pi\ri}$. Both Whittaker functions have branch-cuts (see \textbf{13.14.11} and \textbf{13.14.12} of Ref.~\cite{DLMF}),
\begin{eqnarray}
W_{\kappa,\mu}(z_0\e^{2\pi\ri}) &=& - \e^{-2\pi\ri \mu} W_{\kappa,\mu}(z_0) \nn\\
&&\quad + \frac{2\pi\ri}{\Gamma\left(\half-\mu-\kappa\right)\Gamma(1+2\mu)} M_{\kappa,\mu}(z_0),\nn\\
M_{\kappa,\mu}(z_0\e^{2\pi\ri}) &=&  - \e^{-2\pi\ri \mu} M_{\kappa,\mu}(z_0).
\end{eqnarray}
Using the above explicit results, we find the eigenphases of the Floquet operator $U(T)$ to be the following:
\begin{equation}
U(T) |F^\pm(0)\> = \exp\left(\pm 2\pi \ri \frac{p_3}{\omega}\right) |F^\pm(0)\>.
\label{eqn:65}
\end{equation}
The system is stable, \ie, having two different real eigenphases if the parameter $p_3$ is real and nonzero,
with two different cyclic states explicitly obtained as follows:
\begin{eqnarray}
|F^+(t)\> &=& \left(\begin{array}{c}
M_{\kappa,\mu}(z)\\
\alpha(z) M_{\kappa,\mu}(z) + \beta M_{\kappa,\mu}'(z)
\end{array}\right),\\
|F^-(t)\> &=&\frac{\pi}{\sin(2\pi \mu) \Gamma\left(\half-\mu-\kappa\right)\Gamma(1+2\mu)}  |F^+(t)\> \nn\\
&&\qquad +  \left(\begin{array}{c}
W_{\kappa,\mu}(z)\\
\alpha (z) W_{\kappa,\mu}(z) + \beta W'_{\kappa,\mu}(z)
\end{array}\right).
\end{eqnarray}

A few observations are in order. Firstly, the state $|F^+(t)\>$ only contains one special function $M_{\kappa,\mu}(z)$ and its derivative $M'_{\kappa,\mu}(z)$. {One possibility is that the small-$\omega$ behavior of this cyclic state is governed by one common dominating exponential function. In this case, the adiabatic following dynamics of $|F^+(t)\>$ will be smooth and it can be expected to be close to the smooth behavior of one instantaneous energy eigenstates. Another possibility is that a single special function may still lead to hopping behavior near the Stokes lines, see Sec.~\ref{sec:Airy} for more details.} Secondly and by contrast, the state $|F^-(t)\>$ involves a sum of two different special functions $M_{\kappa,\mu}(z)$ and $W_{\kappa,\mu}(z)$ of different exponential behavior.  Therefore,  due to the Stokes phenomenon, the relative importance of $W_{\kappa,\mu}(z)$ and $M_{\kappa,\mu}(z)$ in this solution can swap during a time window that is rather short as compared with $T$. This being the case, when analyzing $|F^-(t)\>$ using smoothly-changing states such as two instantaneous energy eigenstates, a hopping behavior can emerge. Thus, the Floquet states of this model can behave very similarly with those states shown in Fig.~\ref{fig:UCF} for a different model.  One should not be surprised by such a similarity because the two models are related by a time independent rotation. Note that here we do not need to perform an explicit asymptotic analysis to the Whittaker functions as $\omega\to0$ \cite{footnote3}, considering that Sec.~\ref{sec:exponent} already outlined a general method without referring to exact solutions.

\subsubsection{$\mathbf{p}\cdot\mathbf{q}=0$ and $\mathbf{q}\cdot\mathbf{q}\neq0$}

If one of $\mathbf{p}\cdot\mathbf{q}$ and $\mathbf{q}\cdot\mathbf{q}$ vanishes, but not both, then the solutions are Bessel functions. For example, if $\mathbf{p}\cdot\mathbf{q}=0$ and $\mathbf{q}\cdot\mathbf{q}\neq0$, by the following changing variables,
\begin{equation}
z(t)\equiv \frac{\ri\sqrt{\mathbf{q}\cdot\mathbf{q}}}{\omega} \eiot \quad \mathrm{and} \quad a(z)\equiv a(t),
\label{eqn:68}
\end{equation}
\Eqn{eqn:Whittaker} can be transformed into a Bessel equation (see \textbf{10.13.1} of Ref.~\cite{DLMF}),
\begin{equation}
a''(z) + \left[1 - \frac{p_3(p_3+\omega)}{z^2\omega^2}\right] a(z)=0.
\end{equation}
The solution of $a(t)$ is
\begin{equation}
a(t) = C_1 \sqrt{z} J_\nu\left(z \right) + C_2 \sqrt{z} Y_{\nu}\left(z \right)
\end{equation}
with $
\nu\equiv \frac{1}{2}+\frac{p_3}{\omega}. $
The lower component $b(t)$ is determined by \Eqn{eqn:bt} with
\begin{eqnarray}
\alpha(z) &=& - \frac{q_3}{q_1-\ri q_2} - \ri \frac{p_3}{q_1 - \ri q_2}\frac{\sqrt{\mathbf{q}\cdot\mathbf{q}}}{\omega\, z}, \nn\\
\beta &=& -\ri \frac{\sqrt{\mathbf{q}\cdot\mathbf{q}}}{q_1 - \ri q_2}.
\end{eqnarray}
At $t=T$, $z=z_0\e^{2\pi\ri}$. Both Bessel functions have branch-cuts (see \textbf{10.11.1} and \textbf{10.11.2} of Ref.~\cite{DLMF}),
\begin{eqnarray}
J_{\nu}(z_0\e^{2\pi\ri}) &=& \e^{2\pi\ri \nu} J_{\nu}(z_0),\nn\\
Y_{\nu}(z_0\e^{2\pi\ri}) &=&  \e^{-2\pi\ri \nu} Y_{\nu}(z_0) + 2\ri\cos^2(\nu\pi) J_\nu(z_0).
\end{eqnarray}
The eigenphases of the Floquet operator $U(T)$ are
\begin{equation}
U(T) |F^\pm(0)\> = \exp\left(\pm 2\pi \ri \frac{p_3}{\omega}\right) |F^\pm(0)\>.
\label{eqn:73}
\end{equation}
The system is generically stable (i.e., stable regardless of the actual value of small $\omega$) if the parameter $p_3$ is real and nonzero.

The cyclic states are
\begin{eqnarray}
|F^+(t)\> &=& \left(\begin{array}{c}
2z J_\nu(z)\\
\left[2 z \alpha(z) +\beta \right] J_\nu(z) + 2 z \beta J_\nu'(z)
\end{array}\right),\\
|F^-(t)\> &=&\cos(\nu\pi)  |F^+(t)\> \nn\\
&&\quad  - \sin(\nu\pi) \left(\begin{array}{c}
2z Y_\nu(z)\\
\left[2 z \alpha(z) +\beta\right] Y_\nu(z) + 2 z \beta Y_\nu'(z)
\end{array}\right).\nn\\
\end{eqnarray}
Again, because the state $|F^-(t)\>$ involves a sum of two different special functions, it may show the hopping behavior in the slow-driving limit, very much similar to the Berry-Uzdin model studied earlier \cite{gongwang2018,BU}.

\subsubsection{$\mathbf{p}\cdot\mathbf{q}\neq0$ and $\mathbf{q}\cdot\mathbf{q}=0$}
Similarly, if $\mathbf{q}\cdot\mathbf{q}=0$ and $\mathbf{p}\cdot\mathbf{q}\neq0$, \Eqn{eqn:Whittaker} can be again transformed into a Bessel equation (see \textbf{10.13.5} of Ref.~\cite{DLMF}),
\begin{equation}
z^2 a''(z) -z a'(z) + \left[z^2 - \frac{4p_3(p_3+\omega)}{\omega^2}\right] a(z)=0
\end{equation}
by the following changing variables,
\begin{equation}
z(t)\equiv \frac{2\ri\sqrt{2\mathbf{p}\cdot\mathbf{q}}}{\omega} \e^{\frac{\ri\omega t}{2}}\quad \mathrm{and} \quad a(z)\equiv a(t).
\label{eqn:77}
\end{equation}
the solution of $a(t)$ is
\begin{equation}
a(t) = C_1 z J_\nu\left(z\right) + C_2 z Y_{\nu}\left(z\right)
\end{equation}
with
$
\nu\equiv 1+2\frac{p_3}{\omega}.
$
The lower component $b(t)$ is determined by \Eqn{eqn:bt} with
\begin{eqnarray}
\alpha(z) &=& - \frac{q_3}{q_1-\ri q_2} + \frac{8 p_3}{q_1 - \ri q_2}\frac{\mathbf{p}\cdot\mathbf{q}}{\omega^2 z^2}, \nn\\
\beta(z) &=& \frac{4\mathbf{p}\cdot\mathbf{q}}{(q_1 - \ri q_2)\omega\, z}.
\end{eqnarray}
The eigenphases of the Floquet operator $U(T)$ are
\begin{equation}
U(T) |F^\pm(0)\> = \exp\left(\pm 4\pi \ri \frac{p_3}{\omega}\right) |F^\pm(0)\>.
\label{eqn:80}
\end{equation}
The system is generically stable if the parameter $p_3$ is real and nonzero. {One may observe that the Floquet eigenphases in \Eqn{eqn:80} are simply half of those in Eqs.~(\ref{eqn:65}) and (\ref{eqn:73}). This is because the factor $\e^{\frac{\ri\omega t}{2}}$ is used in changing variables in \Eqn{eqn:77} whereas the factor $\eiot$ is used in deriving Eqs.~(\ref{eqn:60}) and (\ref{eqn:68}).}

The cyclic states are
\begin{eqnarray}
|F^+(t)\> &=& \left(\begin{array}{c}
z J_\nu(z)\\
\left[z \alpha(z) + \beta (z) \right] J_\nu(z) + z \beta(z) J_\nu'(z)
\end{array}\right),\\
|F^-(t)\> &=&\cos(\nu\pi)  |F^+(t)\> - \sin(\nu\pi) \nn\\
&&\quad \times\left(\begin{array}{c}
z Y_\nu(z)\\
\left[ z \alpha(z) +\beta(z) \right] Y_\nu(z) + z \beta(z) Y_\nu'(z)
\end{array}\right).\nn\\
\end{eqnarray}
This case is hence analogous to that in the previous subsection.

\subsubsection{$\mathbf{p}\cdot\mathbf{q}=0$ and $\mathbf{q}\cdot\mathbf{q}=0$}
If both $\mathbf{p}\cdot\mathbf{q}$ and $\mathbf{q}\cdot\mathbf{q}$ vanish then the case becomes trivial. The solutions are simply exponential functions,
\begin{eqnarray}
a(t) &=& C_1 \e^{-\ri p_3 t} + C_2 \e^{\ri\omega t+\ri p_3 t},\nn\\
b(t) &=& -C_1 \frac{q_3}{q_1-\ri q_2} \e^{-\ri p_3 t}  - C_2 \frac{2p_3+\omega+q_3\eiot}{q_1-\ri q_2} \e^{\ri p_3 t} . \nn\\
\end{eqnarray}
The eigenphases of the Floquet operator is $\pm 2\pi \frac{p_3}{\omega}$. The system is generically stable if $p_3$ is real and nonzero. There is not much interesting in this simple case in the slow-driving limit.

\subsection{Solvable case B with $q_1 = \ri q_2$}
\label{sec:SFII}
If  it is more convenient to choose a base such that $q_1 = \ri q_2$, then we get another solvable model with five complex parameters and one frequency. The Hamiltonian has the form
\begin{equation}
H_{\rm 01B}(t) = \left(
\begin{array}{cc}
p_3+q_3\,\eiot & p_1-\ri p_2\\
p_1+\ri p_2 +2 q_1\, \eiot & -p_3-q_3\,\eiot
\end{array}\right).
\label{eqn:H_II}
\end{equation}
In this model, $\mathbf{q}\cdot\mathbf{q} = q_3^2$.

\subsubsection{$q_3\neq0$}
If $q_3\neq0$, then by changing variables,
\begin{equation}
z(t) \equiv 2\frac{q_3}{\omega} \eiot \quad \mathrm{and} \quad
w(z) \equiv z^{-c} \e^{z/2}  a(t),
\end{equation}
with
$
c\equiv \frac{\sqrt{\mathbf{p}\cdot\mathbf{p}}}{\omega},
$
we get a confluent hypergeometric equation (see \textbf{13.2.1} of Ref.~\cite{DLMF}),
\begin{equation}
z w''(z) +\left( 1+ 2 c -z\right) w'(z) -\left(\frac{\mathbf{p}\cdot\mathbf{q}}{\omega q_3} +c\right) w(z) =0.
\end{equation}
The solutions are
\begin{eqnarray}
a(t) &=& \e^{-\half z} z^c \left[C_1 V(z) + C_2 M(z)\right]\\
b(t) &=& -\frac{\omega}{p_1-\ri p_2} \e^{-\half z} z^c \left\{ C_1 \left[ \alpha V(z) + z V'(z) \right] \right.\nn\\
&&\quad\left. +  C_2 \left[ \alpha M(z) + z M'(z) \right]
\right\},
\end{eqnarray}
where we introduced a parameter
\begin{equation}
\alpha \equiv \frac{p_3}{\omega} + c = \frac{p_3 + \sqrt{\mathbf{p}\cdot\mathbf{p}}}{\omega}
\end{equation}
and short-handed notations, $V$ and $M$ for the confluent hypergeometric functions, using notations in Ref.~\cite{DLMF},
\begin{eqnarray}
V(z) &\equiv& U(a,b,z), \\
M(z) &\equiv& \mathbf{M}(a,b,z) = \frac{1}{\Gamma(b)} {}_1 F_1 (a,b,z),
\end{eqnarray}
with the parameters
\begin{eqnarray}
a &\equiv& \frac{\mathbf{p}\cdot\mathbf{q}}{\omega q_3} +c = \frac{1}{\omega} \left(  \frac{\mathbf{p}\cdot\mathbf{q}}{q_3} + \sqrt{\mathbf{p}\cdot\mathbf{p}}\right), \\
b &\equiv& 1+2c = 1+ 2 \frac{\sqrt{\mathbf{p}\cdot\mathbf{p}}}{\omega}.
\end{eqnarray}
At $t=T$, $z=z_0\e^{2\pi\ri}$, $z^c$ acquires a phase,
$
z^c = z_0^c \e^{2\pi\ri c}.
$
$\mathbf{M}(a,b,z)$ is an entire function,
$
\mathbf{M}(a,b,z_0 \e^{2\pi\ri}) = \mathbf{M}(a,b,z_0).
$
But $U(a,b,z)$ has a branch-cut,
\begin{equation}
U(a,b,z_0\e^{2\pi\ri}) = \e^{-2\pi\ri b} U(a,b,z_0) + \frac{2\pi\ri \e^{-\pi\ri b}}{\Gamma(1+a-b)} \mathbf{M}(a,b,z_0).
\end{equation}
After plugging parameters into  this model, we have
\begin{equation}
V(z_0\e^{2\pi\ri}) = \e^{-4\pi\ri c} V(z_0) - \frac{2\pi\ri \e^{-2\pi\ri c}}{\Gamma\left(a - 2c \right)} M(z_0).
\end{equation}
The eigenphases of the Floquet operator are then found from
\begin{equation}
U(T) |F^\pm(0)\> =\e^{\pm 2\pi \ri c} |F^\pm(0)\>.
\end{equation}
The system is clearly stable for real and nonvanishing $c$, with one cyclic state given by
\begin{equation}
|F^+(t)\> = z^c \e^{-\half z} \left(\begin{array}{c}
-\left(p_1 -\ri p_2\right) M(z)\\
\omega \left[\alpha M(z) + z M'(z)\right]
\end{array}\right),
\end{equation}
and the other cyclic state given by
\begin{eqnarray}
|F^-(t)\> &=& \frac{\pi}{\sin(2\pi c) \Gamma \left(a-2c \right)}|F^+(t)\> \nn\\
&&\quad +   z^c \e^{-\half z}  \left(\begin{array}{c}
-\left(p_1 -\ri p_2\right) V(z)\\
\omega \left[\alpha V(z) + z V'(z)\right]
\end{array}\right) .
\end{eqnarray}
In the slow-driving limit, we again expect $|F^-(t)\>$ (but not $|F^+(t)\>$) to  hop between two instantaneous energy eigenstates. The hopping occurs when the relative importance of $V(z)$ and $M(z)$ swaps due to the Stokes phenomenon.  In principle we could perform an asymptotic analysis to the confluent hypergeometric functions as $\omega\to0$. But again, it is much easier to analyze the differential equation directly as we did in Sec. \ref{sec:exponent}.

Before ending this subsection, we would like to mention an interesting model studied by a group from the University of Central Florida (UCF) \cite{UCF}. It is actually a special case here with
\begin{eqnarray}
&& p_1= -1, \quad p_2 = 0, \quad p_3=\ri r,\nn\\
&& q_1= 0,\quad q_2=0,\quad q_3 =- \ri\rho.
\end{eqnarray}
With these choice of the parameters, the Hamiltonian has the form
\begin{equation}
H_{\rm UCF} = \left(
\begin{array}{cc}
\ri r-\ri\rho\,\eiot & -1\\
-1 & -\ri r+\ri\rho\,\eiot
\end{array}\right).
\label{eqn:HUCF}
\end{equation}
{For convenience, we rename the parameter $g_0$ in Ref.~\cite{UCF} as $r$, and their $\gamma$ as $\omega$.} In this case,
$
\mathbf{p}\cdot\mathbf{p} = 1-r^2$, $\mathbf{p}\cdot\mathbf{q} = r\rho$, $ \mathbf{q}\cdot\mathbf{q} = -\rho^2.
$

Both absolute adiabatic following and piecewise adiabatic following with hopping are possible in this model, as shown in Fig.~\ref{fig:UCF}. The real part of the critical exponent is also presented in Fig.~\ref{fig:UCF_g}.  Remarkably, it is seen that the hopping timing is fully consistent with the locations where $\Re[g(\omega t)]$ changes its sign.

The model in \Eqn{eqn:HUCF} is parameterized by two complex parameters, $r$ and $\rho$. Each parameter is determined by two real numbers. Therefore, the parameter space of the model has four dimensions. In this four dimensional space, there exists a critical surface. On one side of it, both Floquet states always follow the instantaneous energy eigenstates in the adiabatic limit, \ie, the central clear region in the left panel of Fig.~\ref{fig:rho}. On the other side of it, one Floquet state hops as $\omega\to0$, for example, the shaded regions in Fig.~\ref{fig:rho}. To illustrate, in Fig.~\ref{fig:rho} we plot the phase diagram of the hopping behavior, \ie, the intersection of the critical surface on the $\Re[r]$-$\Re[\rho]$ plane. In the hopping region, there are two possibilities for this model. The cyclic state may hop twice in one period (light-shaded), or it may hop four times (dark-shaded).

{Figure \ref{fig:theta} depicts the timings of the hopping, as determined from $\Re[g(\theta_\textrm{crit})]=0$ for $\rho=\rho_\textrm{crit}$. For {a real $r$ with its magnitude} $|r|\lesssim 0.361$, there are two $\theta_\textrm{crit}$ for each $\rho_\textrm{crit}$, which means that hopping four times within one period is possible. For $0.361\lesssim |r| < 1$, only hopping twice is possible. For  {a real and large $r$ with} $|r|>1$, the system is unstable because $c=\sqrt{1-r^2}$ is complex. }

\begin{figure}[h!]
\begin{center}
\includegraphics[width=0.465\columnwidth]{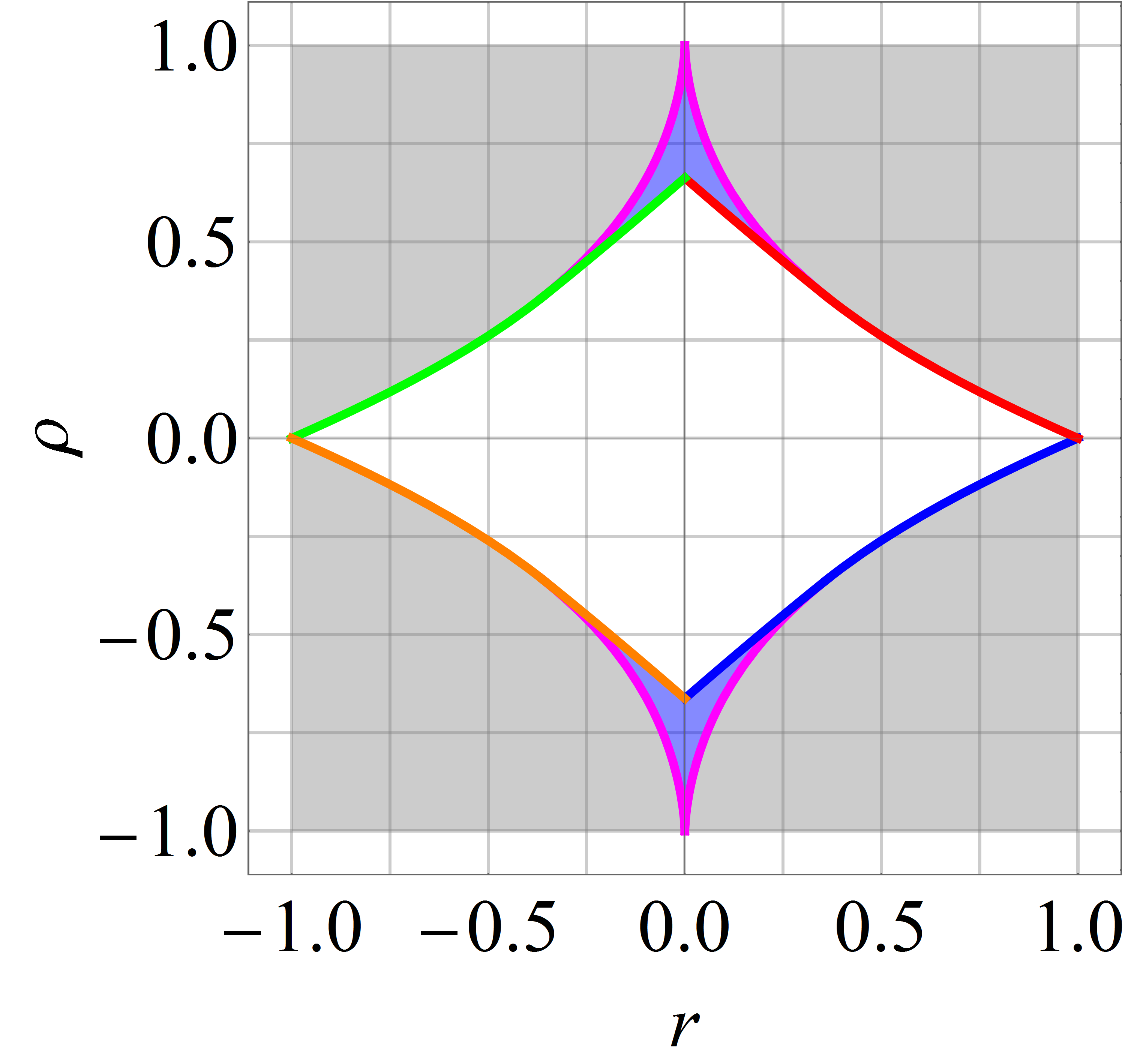}
\includegraphics[width=0.45\columnwidth]{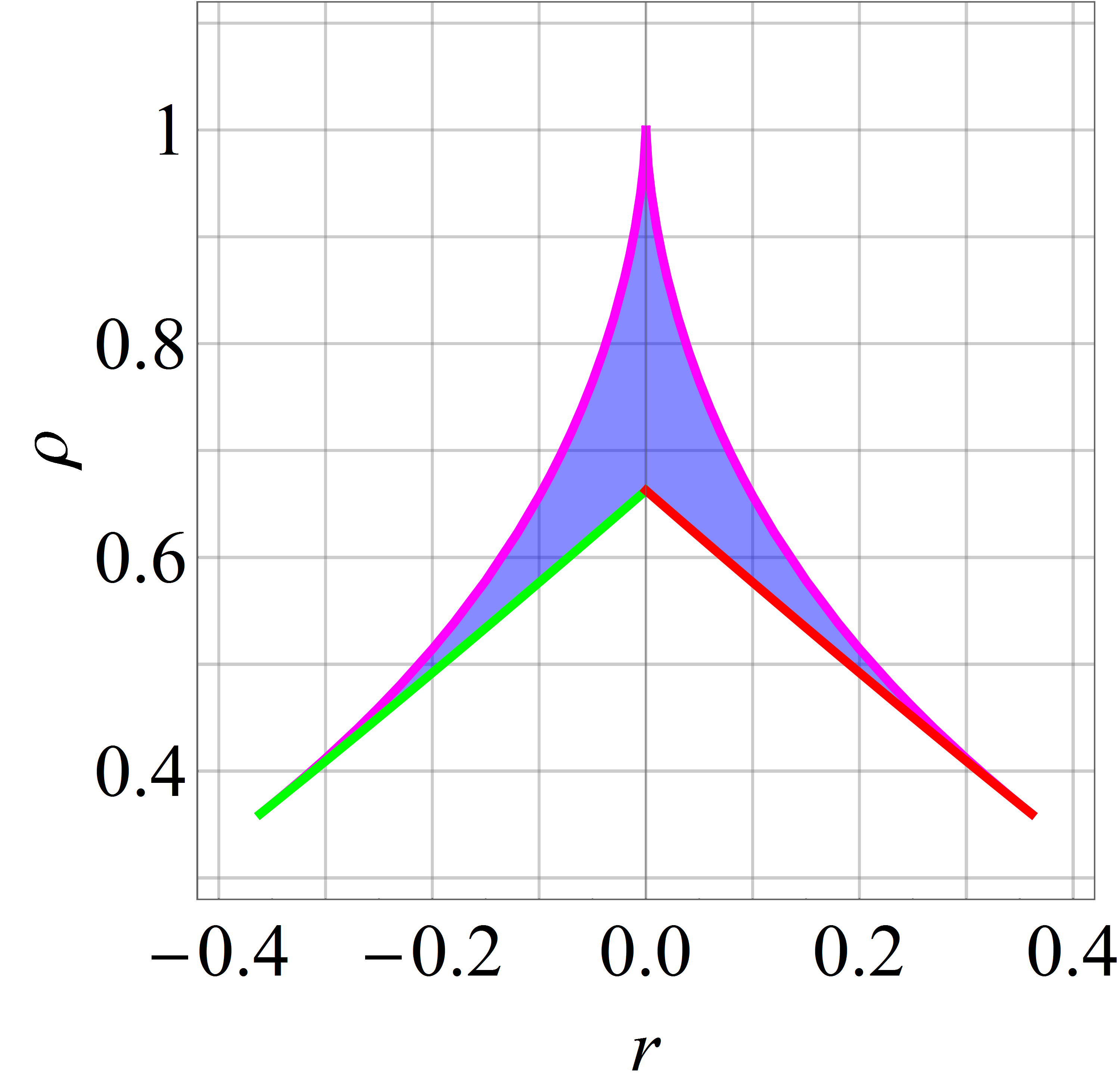}
	\caption{\label{fig:rho}(color online)
	Phase diagram of the hopping behavior on the $\Re[r]$-$\Re[\rho]$ plane in the model depicted by \Eqn{eqn:HUCF}. In the non-shaded area, both cyclic states follow instantaneous energy eigenstates in the adiabatic limit. In the shaded area, one cyclic state exhibits piecewise following in the slow driving limit. States in the light-shaded (gray) region may hop twice in one period, whereas states in the dark-shaded (blue) region may hop four times. The right panel is a zoom-in view for the top region in the left panel.}
\end{center}
\end{figure}

\begin{figure}[h!]
\begin{center}
	\includegraphics[width=0.45\columnwidth]{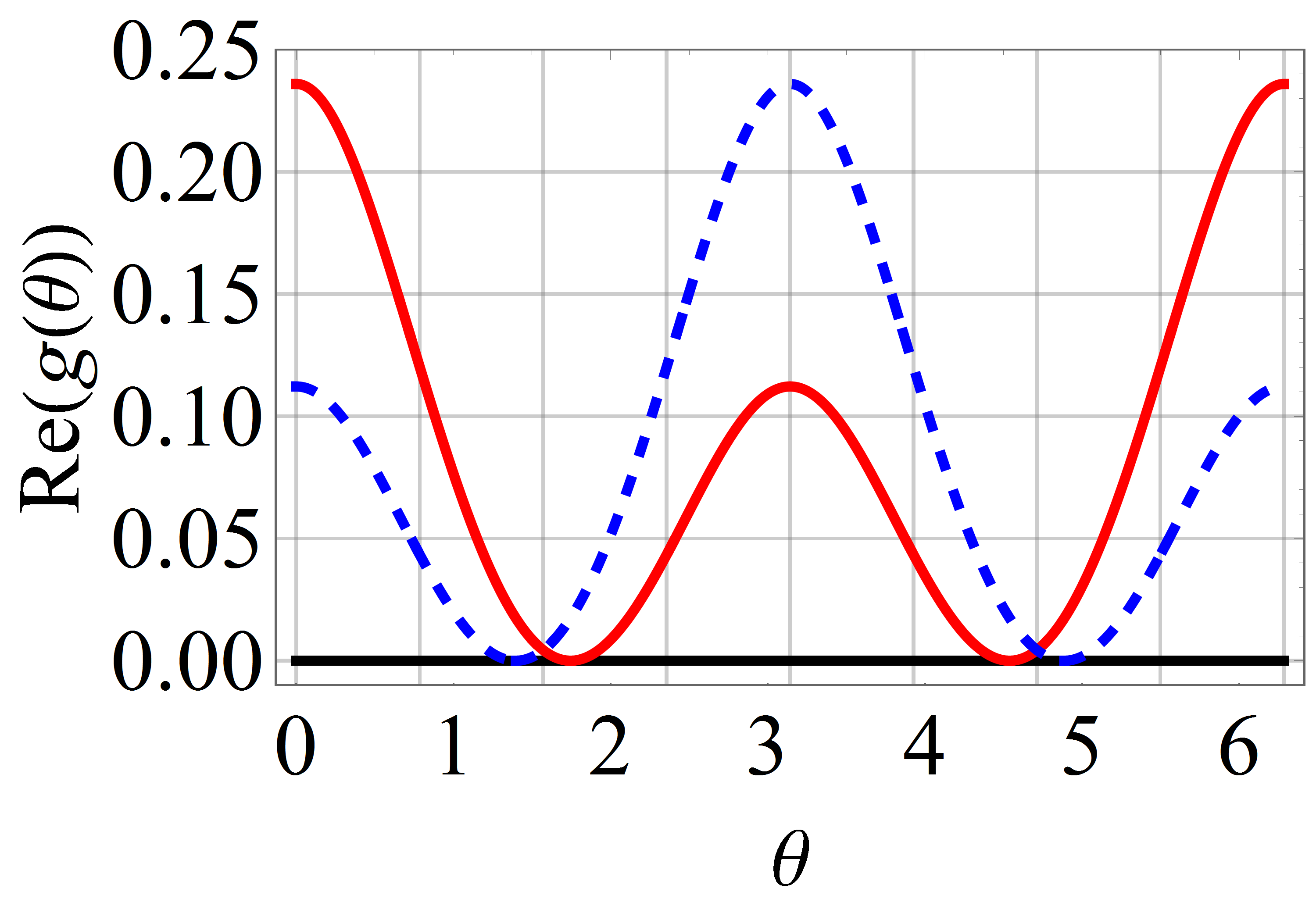}
	\includegraphics[width=0.45\columnwidth]{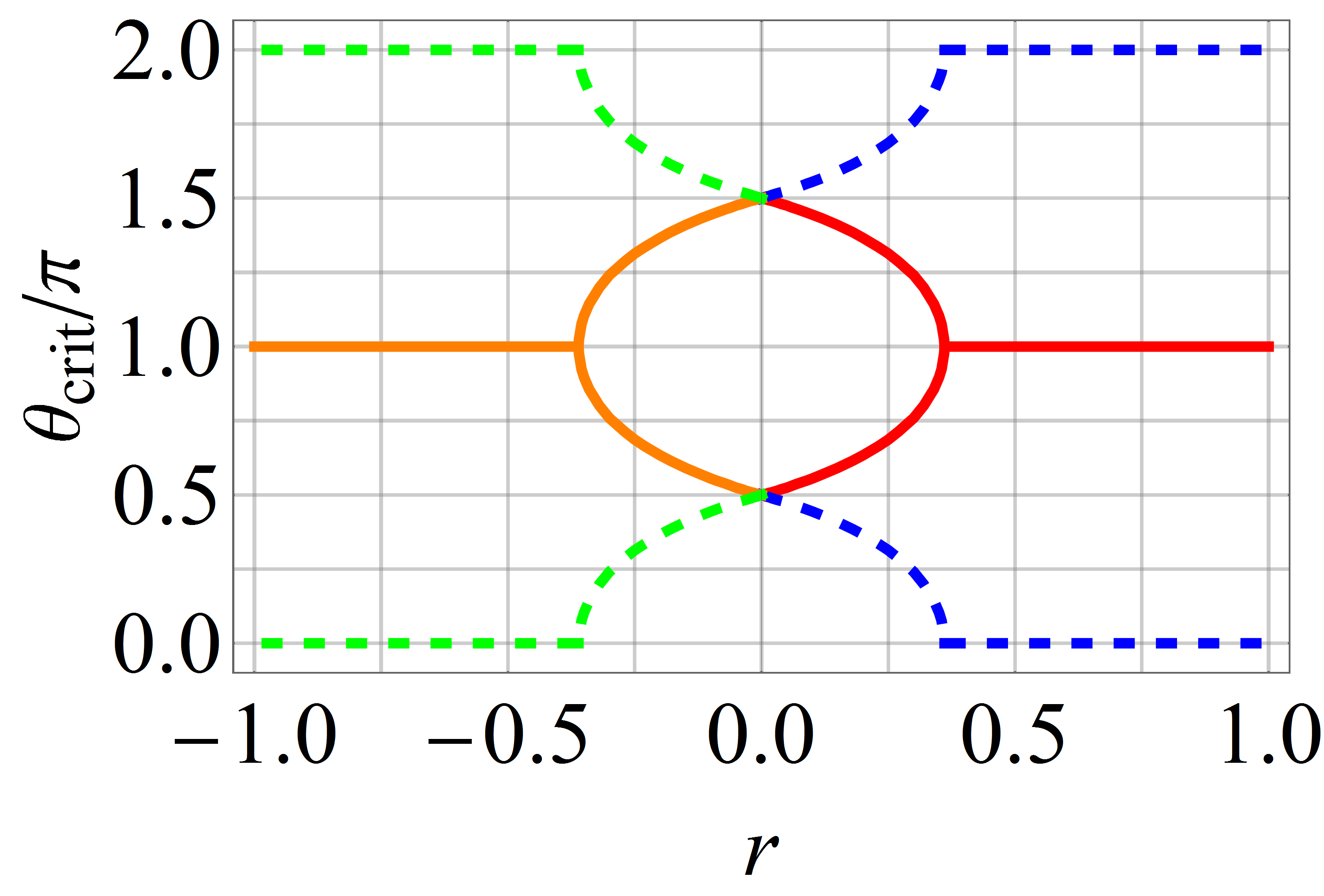}
	\caption{\label{fig:theta}(color online)
	Left panel: Real part of the critical exponent $g$ as a function of $\theta=\omega t$ when $\rho=\rho_\textrm{crit}$ in the model in \Eqn{eqn:HUCF}. In this plot we choose $r=0.1$. $\theta_\textrm{crit}$ are defined as the locations where the curves are tangent to the horizontal axis. From the plot, we can obtain that $\theta_\textrm{crit}=1.395884, 4.886075$ for $\rho_\textrm{crit}= 0.5766416$ [the (red) solid line] and  $\theta_\textrm{crit}=1.744482, 4.538703$ for $\rho_\textrm{crit}= - 0.5766416$ [the (blue) dashed line]. $\theta_\textrm{crit}$ for the case that hops four times is not plotted because it is always at $\pi$.  Right panel: $\theta_\textrm{crit}$ as a function of {a real} $r$ in the same model. (Colors of lines in the two panels here match those in Fig.~\ref{fig:rho}.) The bifurcation/merging of $\theta_\textrm{crit}$ occurs around $r\approx \pm 0.361$.}
\end{center}
\end{figure}

\subsection{$q_3=0$}
If $q_3=0$,
$
\mathbf{p}\cdot\mathbf{q} = \left(p_1-\ri p_2\right)q_1.
$
Further assuming that $\mathbf{p}\cdot\mathbf{q}\neq0$, then \Eqn{eqn:masterz} can be transformed into a Bessel equation. The solutions of $a(t)$ is
\begin{equation}
a(t) = C_1 J_\nu\left(\frac{2\ri\sqrt{2\mathbf{p}\cdot\mathbf{q}}}{\omega} \e^{\frac{\ri \omega t}{2}}\right) + C_2 Y_{\nu}\left(\frac{2\ri\sqrt{2\mathbf{p}\cdot\mathbf{q}}}{\omega} \e^{\frac{\ri \omega t}{2}}\right)
\end{equation}
with
$
\nu\equiv \frac{2\sqrt{\mathbf{p}\cdot\mathbf{p}}}{\omega}.
$
The lower component $b(t)$ is determined by \Eqn{eqn:bt}.

If both $q_3=0$ and $\mathbf{p}\cdot\mathbf{q}=0$, then $p_1=\ri p_2$. {Combining with the choice $q_1=\ri q_2$ in this subsection, we get a lower triangular form of matrix Hamiltonian. It is better studied by solving $b(t)$ first, which will not be elaborated here.}

\subsubsection{Berry-Uzdin model}
The Berry-Uzdin model is a special case with \cite{BU}
\begin{eqnarray}
&& p_1= \half \ri (1-r), \quad p_2 = -\half(1+r), \quad p_3=0,\nn\\
&& q_1=\half\ri\rho,\quad q_2=\half\rho,\quad q_3 =0.
\end{eqnarray}
With these choice of the parameters, the Hamiltonian has the form
\begin{equation}
H_{\rm BU} =\ri \left(
\begin{array}{cc}
0&1\\
-r+\rho\,\eiot&0
\end{array}\right).
\end{equation}
In this case,
$
\mathbf{p}\cdot\mathbf{p} = r$, $\mathbf{p}\cdot\mathbf{q} = -\half\rho$, $ \mathbf{q}\cdot\mathbf{q} = 0.
$
The differential equation in Eq.~(\ref{eqn:masterz}) becomes
\begin{equation}
\omega^2 Z^2 a''(Z) + \omega^2 Z a'(Z) -(r-\rho Z)a(Z)=0,
\end{equation}
which can be converted to a Bessel equation. The solutions are
\begin{equation}
a(Z) = C_1 J_\nu\left(2\frac{\sqrt{\rho Z}}{\omega}\right) + C_2 Y_{\nu}\left(2\frac{\sqrt{\rho Z}}{\omega}\right)
\end{equation}
with
$
\nu\equiv 2\frac{\sqrt{r}}{\omega}.
$
In term of the original variable,
\begin{equation}
a(t) = C_1 J_\nu\left(2\frac{\sqrt{\rho}}{\omega}\e^{\frac{\ri \omega t}{2}}\right) + C_2 Y_{\nu}\left(2\frac{\sqrt{\rho}}{\omega}\e^{\frac{\ri \omega t}{2}}\right),
\end{equation}
and
$
b(t) = \dot{a}(t).
$
The constants $C_1$ and $C_2$ are determined by the initial state. We do not discuss this model further as it was studied in detail in our previous work \cite{gongwang2018}.

\section{Two-frequency models}
In this section, we consider models with two Fourier components,
\begin{equation}
\mathbf{f}(t) = \mathbf{p} \e^{\ri m \omega t} + \mathbf{q}\e^{\ri n \omega t},
\end{equation}
with $m\neq0$ and $n\neq0$. The Hamiltonian
\begin{equation}
H_{mn} = \mathbf{p}\cdot\bm{\sigma} \e^{\ri m \omega t} + \mathbf{q}\cdot\bm{\sigma} \e^{\ri n \omega t},
\end{equation}
We extend the technique used in the first part of this work by first changing variables,
$
Z\equiv \eiot$ and $a(Z) = a(t)$.
We get a more complicated ``master equation" in variable $Z$ as
\begin{equation}
(p_1-\ri p_2)Z^m \{\mathrm{eqn}_P\} + (q_1-\ri q_2) Z^n  \{\mathrm{eqn}_Q\} = 0,
\label{eqn:masterz12}
\end{equation}
with
\begin{eqnarray}
\mathrm{eqn}_P &=& \omega^2 Z^2 a''(Z) - \omega^2 (m-1) Z a'(Z) \nn\\
&& \quad - \left[\mathbf{p}\cdot\mathbf{p} Z^{2m} + 2 \mathbf{p}\cdot\mathbf{q} Z^{m+n} + \mathbf{q}\cdot\mathbf{q} Z^{2n} \right.\nn\\
&&\quad\left. + \omega(m-n)q_3 Z^n \right] a(Z), \nn\\
\mathrm{eqn}_Q &=& \omega^2 Z^2 a''(Z) - \omega^2 (n-1) Z a'(Z) \nn\\
&& \quad - \left[\mathbf{p}\cdot\mathbf{p} Z^{2m} + 2 \mathbf{p}\cdot\mathbf{q} Z^{m+n} + \mathbf{q}\cdot\mathbf{q} Z^{2n} \right.\nn\\
&&\quad\left.+ \omega(n-m) p_3 Z^m \right] a(Z). \nn
\end{eqnarray}
Again, let
$
a(Z) \equiv \e^{\frac{m(Z)}{\omega}},
$
then $m(Z)$ is found to have a smooth slow driving limit. To the leading order, we arrive at
\begin{equation}
m(Z)\sim \pm\int^Z\rd x\frac{\sqrt{\mathbf{p}\cdot\mathbf{p}\, x^{2m} + 2\mathbf{p}\cdot\mathbf{q}\,x^{m+n} +\mathbf{q}\cdot\mathbf{q}\,x^{2n}}}{x}.
\label{eqn:m(z)12}
\end{equation}
In general, neither the ``master" equation in Eq.~(\ref{eqn:masterz12}) nor this critical exponent is analytically solvable.

To look into some special solvable cases, we may rescale $\omega\to\omega/m$ to absorb the parameter $m$. Thus, without loss of generality, let us consider
\begin{equation}
H_{1r} = \mathbf{p}\cdot\bm{\sigma} \e^{\ri \omega t} + \mathbf{q}\cdot\bm{\sigma} \e^{\ri r \omega t},
\end{equation}
where $r=n/m$ is a rational number.

\subsection{Solvable model with $r=2$}

The simplest Hamiltonian $H_{1r}$ is the case with $r=2$.
\begin{eqnarray}
H_{12} &=& \mathbf{p}\cdot\bm{\sigma} \eiot + \mathbf{q}\cdot\bm{\sigma} \e^{2\ri \omega t}\nn\\
&=& \left(
\begin{array}{cc}
p_3 		&	p_1-\ri p_2\\
p_1 +\ri p_2 	&	-p_3
\end{array}
\right) \eiot \nn\\
&&\quad + \left(
\begin{array}{cc}
q_3 		&	q_1-\ri q_2\\
q_1 + \ri q_2 	&	- q_3
\end{array}
\right) \e^{2 \ri \omega t}.
\label{eqn:H12}
\end{eqnarray}
For this model, the critical exponent in \Eqn{eqn:m(z)12} has a closed form. Using the integration formula \textbf{2.261} and \textbf{2.262} in Ref.~\cite{RG},
\begin{eqnarray}
&&\int\rd x \sqrt{a + b x +cx^2} \nn\\
&=& \frac{b+ 2c x}{4c }\sqrt{a + bx +cx^2} \nn\\
&&\quad - \frac{b^2-4ac}{16c^{3/2}} \ln\frac{b + 2c x + 2\sqrt{c} \sqrt{a + bx +cx^2}}{b+2cx - 2\sqrt{c} \sqrt{a + bx +cx^2}}. \nonumber \\
\label{eqn:integral2}
\end{eqnarray}
The integration constant is chosen such that the integral vanishes when $\sqrt{a + bx +cx^2}=0$. Just like before,
we define the exponent function $g(\omega t)$ to be the right hand side of Eq.~(\ref{eqn:integral2}) with $a= \mathbf{p}\cdot\mathbf{p}$, $ b= 2\mathbf{p}\cdot\mathbf{q}$, $c=\mathbf{q}\cdot\mathbf{q}$, and $x=\eiot$. A sudden-switch behavior is then expected in the adiabatic following dynamics if the real part of $g(\omega t)$ obtained above
flips signs during the time evolution.

The model $H_{12}$ becomes exactly solvable if one rotates the coordinates such that $q_1=\ri q_2$. As we argued earlier, this is always doable. In this new coordinate system, $\mathbf{q}\cdot\mathbf{q} = q_3^2$.

\subsubsection{$q_3 \neq 0$.}

If $q_3 \neq 0$, the solutions are parabolic cylinder functions. To see this, let us change variables
\begin{equation}
z(t) \equiv \sqrt{\frac{2}{\omega}} \left(\frac{\mathbf{p}\cdot\mathbf{q}}{q_3^{3/2}} + \sqrt{q_3} \eiot \right), \qquad a(z) \equiv a(t).
\end{equation}
Then $a(z)$ satisfies a parabolic cylinder equation,
\begin{equation}
a''(z) + \left(\nu+\half - \quarter z^2\right) a(z)=0,
\end{equation}
with
$
\nu \equiv  \frac{(\mathbf{p}\cdot\mathbf{q})^2}{2\omega q_3^3} - \frac{ \mathbf{p}\cdot\mathbf{p} }{2\omega q_3}.
$
The solutions are
\begin{eqnarray}
a(z) &=& C_1 D_\nu(z) + C_2 D_\nu(-z), \nn\\
b(z) &=& \alpha(z) a(z) + \beta a'(z),
\label{eqn:Dnu}
\end{eqnarray}
with
\begin{eqnarray}
\alpha(z) &=& \frac{q_1}{q_3} - \frac{z}{p_1-\ri p_2} \sqrt{\frac{\omega q_3}{2}}, \nn\\
\beta &\equiv& - \frac{\sqrt{2\omega q_3}}{p_1-\ri p_2}.
\end{eqnarray}

Because $D_\nu(z)$ and $\alpha(z)$ are entire functions in the complex $z$-plane, the time evolution is periodic. That is
$
U(T) = U(0) = 1.
$
The eigenvalue of the Floquet operator is simply unity and any state would be cyclic in this model. The system is stable for arbitrary choice of parameters. However, piecewise adiabatic following still presents, \ie, if the real part of the critical exponent $g(\theta)$ flips its sign due to the underlying Stokes phenomenon.

\subsubsection{$q_3=0$}
\label{sec:Airy}
The conditions $q_3 = 0$ and $q_1=\ri q_2$ means that $\mathbf{q}\cdot\mathbf{q}=0$.  {The Hamiltonian is parametrized by four complex parameters,
\begin{eqnarray}
H_\textrm{Airy} &=& \mathbf{p}\cdot\bm{\sigma} \eiot + \mathbf{q}\cdot\bm{\sigma} \e^{2\ri \omega t}\nn\\
&=& \left(
\begin{array}{cc}
p_3 		&	p_1-\ri p_2\\
p_1 +\ri p_2 	&	-p_3
\end{array}
\right) \eiot + \left(
\begin{array}{cc}
0 		&	0\\
2q_1 	&	0
\end{array}
\right) \e^{2 \ri \omega t}.
\label{eqn:HAiry} \nn\\
\end{eqnarray}
}
In this case, the solutions are Airy functions. To see this, let us change variables
\begin{eqnarray}
z(t) &\equiv& \frac{1}{\omega^{2/3}} \left[ \frac{\mathbf{p}\cdot\mathbf{p}}{(2\mathbf{p}\cdot\mathbf{q})^{2/3}} + \left(2 \mathbf{p}\cdot\mathbf{q}\right)^{1/3} \eiot \right], \nn\\
a(z) &\equiv& a(t).
\end{eqnarray}
Then $a(z)$ satisfies the Airy equation,
\begin{equation}
a''(z) =z a(z).
\end{equation}
The solutions are
\begin{eqnarray}
a(z) &=& C_1 \mathrm{Ai}(z) + C_2 \mathrm{Bi}(z), \nn\\
b(z) &=& \alpha a(z) + \beta a'(z),
\end{eqnarray}
with
\begin{equation}
\alpha = -\frac{p_3}{p_1-\ri p_2} \quad \mathrm{and} \quad \beta \equiv - \frac{(2\omega q_1)^{1/3}}{(p_1-\ri p_2)^{2/3}}.
\end{equation}

Because $\mathrm{Ai}(z)$ and $\mathrm{Bi}(z)$ are entire functions in the complex $z$-plane, the time evolution is periodic. That is
$
U(T) = U(0) = 1.
$
The eigenvalue of the Floquet operator is simply unity again.  So any state would be cyclic and stable. Consider next specifically the critical exponent $g(\theta)$:
\begin{eqnarray}
g(\theta) &=& \int^Z \rd\, x \sqrt{\mathbf{p}\cdot\mathbf{p} + 2 \mathbf{p}\cdot\mathbf{q}\, x}\nn\\
&=& \frac{1}{3 \mathbf{p}\cdot\mathbf{q}} \left( \mathbf{p}\cdot\mathbf{p} + 2 \mathbf{p}\cdot\mathbf{q} \e^{\ri\theta}\right)^{3/2}.
\end{eqnarray}

{We now exploit this model to elaborate how an expected sudden-switch in the adiabatic following dynamics may be suppressed.  As shown in Figs.~\ref{fig:Ai_g} and \ref{fig:Airy}, the critical exponent $\Re[g(\theta)]$ changes its sign three times within one driving period. As expected, the solution with $a(z)=\mathrm{Bi}(z)$ the (red) dot-dashed lines in Fig.~\ref{fig:Airy}] hops every time when $\Re[g(\theta)]=0$. However, the solution with $a(z)=\mathrm{Ai}(z)$ [the (purple) solid lines in Fig.~\ref{fig:Airy}] only hops \textit{once} around $t=T/2$. To develop more understandings, note that the Airy function $\mathrm{Ai}(z)$ is also the so-called recessive (subdominant) solution, whose asymptotic expansion has only one term with a negative real part in the critical exponent for positive real $z$. That is, in that regime there is no dominating exponential term \cite{BenderBook}. Consider next what happens after crossing the Stokes line at $\arg(z)=\frac{\pi}{3}$, \ie, where $\Re[g(\theta)]=0$. Then the real part of the critical exponent of $\mathrm{Ai}(z)$ changes from negative to positive. That is, a dominant exponential term emerges only after crossing the Stokes line.  Since there is no switch from one dominating exponential term to a different one upon crossing the Stokes line, such a recessive solution does not display any sudden change when projected onto smooth basis states.   Therefore, no hopping at $\arg(z)=\frac{\pi}{3}$. By symmetry, there is no hopping for $\mathrm{Ai}(z)$ at  $\arg(z)=-\frac{\pi}{3}$ either. Finally, at the next Stokes line at  $\arg(z)=\pi$, a previously subdominant (dominant) term becomes dominant (subdominant), and a sudden-switch behavior emerges.  By contrast,  for a dominant solution like $\mathrm{Bi}(z)$, as soon as the first Stokes line at $\arg(z)=\frac{\pi}{3}$ is crossed, the previously dominant (subdominant) term becomes subdominant (dominant), and therefore hopping occurs there. To end the discussions here, we provide the asymptotic expansions of $\mathrm{Ai}(z)$ and $\mathrm{Bi}(z)$ in connection with their respective Stokes lines:
\begin{equation}
\mathrm{Ai}(z)\sim \frac{1}{2\sqrt{\pi}z^{1/4}} \e^{-\frac{2}{3}z^{3/2}}, \qquad |z|\to\infty
\end{equation}	
is valid for $|\arg(z)|<\pi$, but
\begin{equation}
\mathrm{Bi}(z)\sim \frac{1}{\sqrt{\pi}z^{1/4}} \e^{+\frac{2}{3}z^{3/2}}, \qquad |z|\to\infty
\end{equation}
is only valid for $|\arg(z)|<\frac{\pi}{3}$ \cite{BenderBook}.}

\begin{figure}[h!]
	\begin{center}
		\includegraphics[width=0.52\columnwidth]{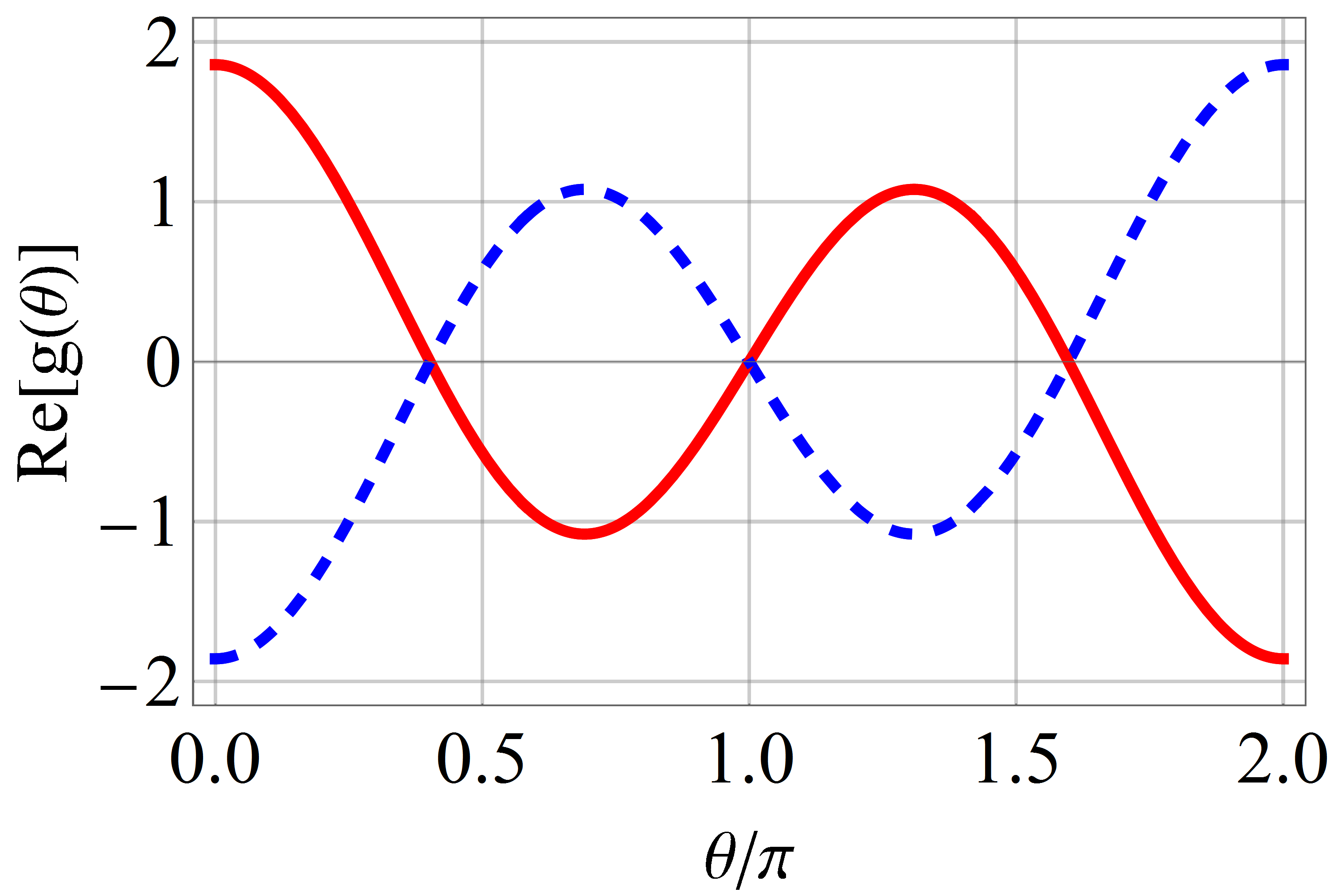}
		\includegraphics[width=0.44\columnwidth]{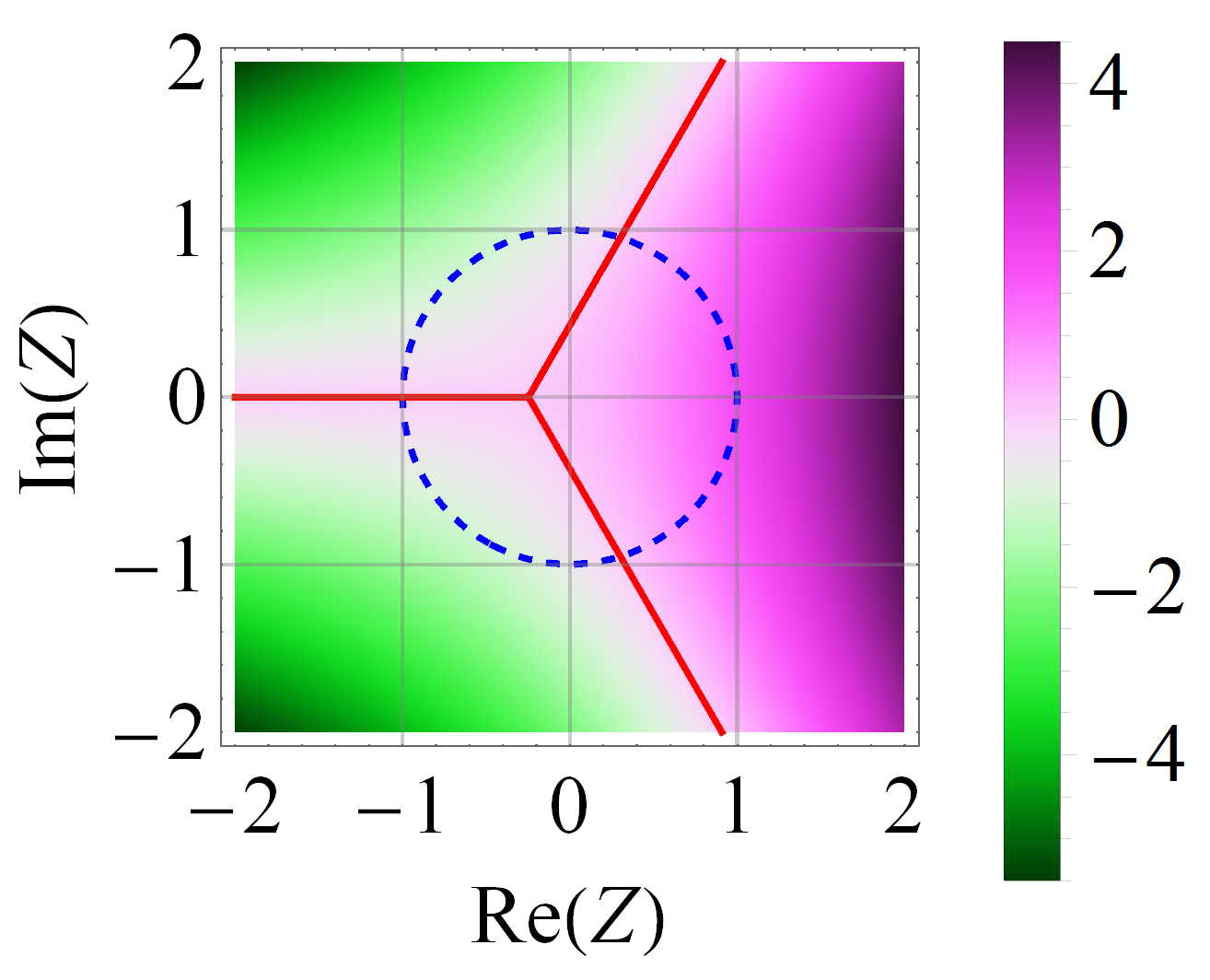}
		\caption{\label{fig:Ai_g}(color online)
			Left panel: Real part of the critical exponent $g$ as a function of $\theta$ in the model depicted in \Eqn{eqn:HAiry}. The parameters are $p_1=1$, $p_2=0$, $p_3=0.5\ri$, and $q_1=0.6$. The (blue) dashed line corresponds to $a(z) = \mathrm{Ai}(z)$ and the (red) solid line corresponds to $a(z) = \mathrm{Bi}(z)$. Right panel: Density plot of $\Re[g(\theta)]$. The (red) solid lines are the Stokes lines, $\arg(z)=\pm\frac{\pi}{3},~\pi$. The (blue) dashed line is a unit circle in $Z$. The hopping occurs when the two lines intersect for the solution $\mathrm{Bi}(z)$, whereas the solution $\mathrm{Ai}(z)$ only hops once at the intersection with $\arg(z)=\pi$.}
	\end{center}
\end{figure}

\begin{figure}[h!]
	\begin{center}
		\includegraphics[width=0.48\columnwidth]{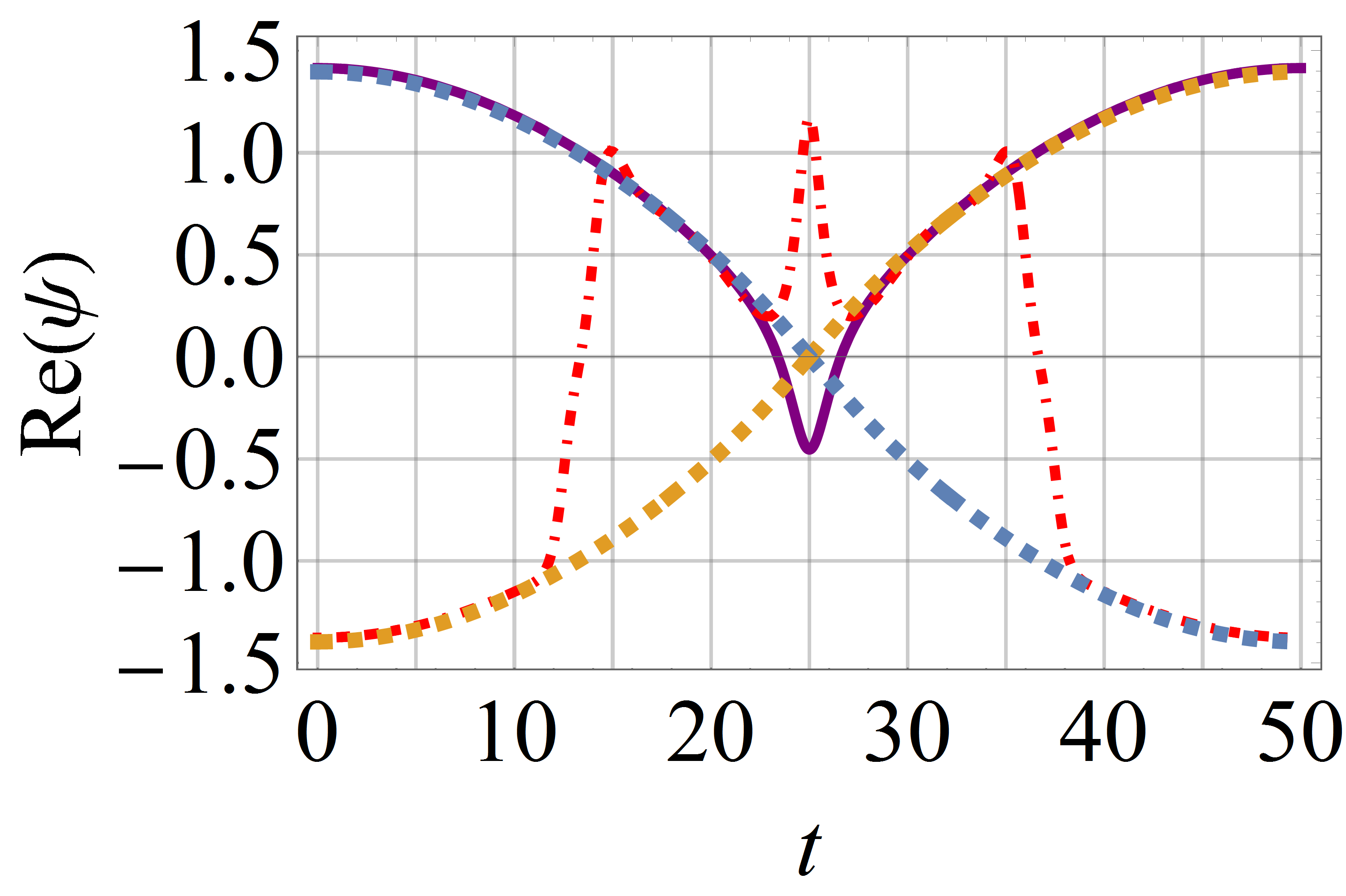}
		\includegraphics[width=0.48\columnwidth]{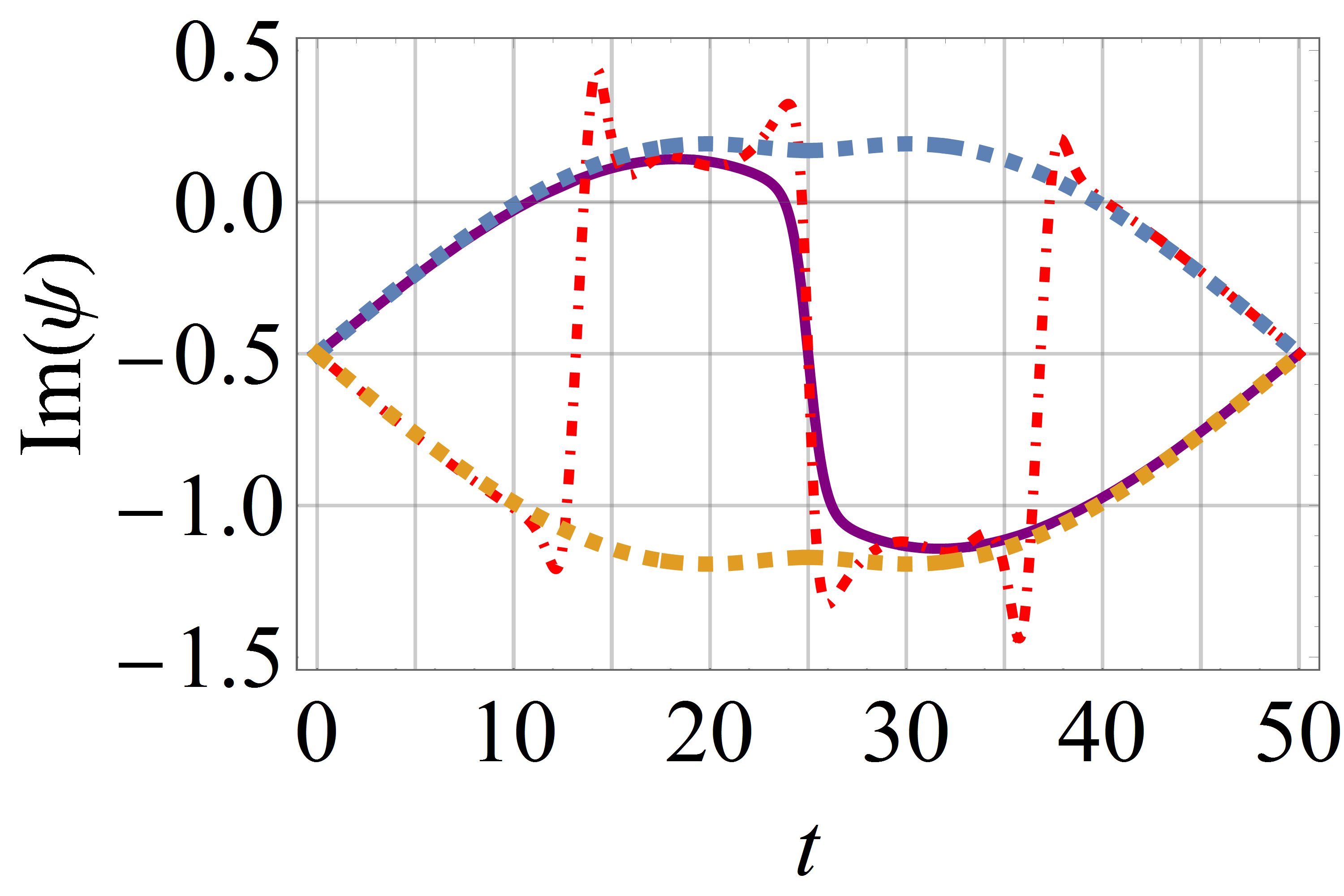}
		\caption{\label{fig:Airy}(color online)
			Real and imaginary parts of $\psi=b(t)/a(t)$ during the time evolution of cyclic states (solid and dot-dashed lines) and of the instantaneous energy eigenstates (dashed lines) for the model depicted in Fig.~\ref{fig:Ai_g} with $\omega = 2\pi/50$. Note that the recessive solution $a(z)=\textrm{Ai}(z)$ [(purple) solid lines] hops only near $t=T/2$, but the dominant solution $a(z)=\textrm{Bi}(z)$ [(red) dot-dashed lines] hops three times within one period.}
	\end{center}
\end{figure}

\subsection{Solvable model with $r=-1$}

For completeness let us consider the Hamiltonians $H_{1r}$ with $r=-1$. For convenience, we put $\mathbf{p}$ in front of the lower frequency term. That is,
\begin{eqnarray}
H_{1-1} &=& \mathbf{p}\cdot\bm{\sigma} \e^{-\ri \omega t} + \mathbf{q}\cdot\bm{\sigma} \eiot\nn\\
&=& \left(
\begin{array}{cc}
p_3 		&	p_1-\ri p_2\\
p_1 +\ri p_2 	&	-p_3
\end{array}
\right) \e^{- \ri \omega t} \nn\\
&&\quad + \left(
\begin{array}{cc}
q_3 		&	q_1-\ri q_2\\
q_1 + \ri q_2 	&	- q_3
\end{array}
\right) \e^{ \ri \omega t}.
\end{eqnarray}
This Hamiltonian has a symmetry under $\mathbf{p}\leftrightarrow\mathbf{q}$ and $\omega\leftrightarrow-\omega$. Thus all the following discussion in this subsection can be easily extended to $\mathbf{p}\leftrightarrow\mathbf{q}$ and $Z\leftrightarrow1/Z$.

For this model, the critical exponent in \Eqn{eqn:m(z)12} has the form,
\begin{equation}
m(Z)\sim \pm\int^Z\rd x\frac{\sqrt{\mathbf{p}\cdot\mathbf{p} + 2\mathbf{p}\cdot\mathbf{q}\,x^2 +\mathbf{q}\cdot\mathbf{q}\,x^4}}{x^2},
\label{gc52}
\end{equation}

For the model $H_{1-1}$ to be solvable, either $\mathbf{p}\cdot\mathbf{p}=0$ or $\mathbf{q}\cdot\mathbf{q}=0$. Since there is the internal symmetry between two frequency components, we consider $\mathbf{p}\cdot\mathbf{p}=0$ without loss of generality. If we separate $\mathbf{p}$ into its real and imaginary parts as
$
\mathbf{p} = \mathbf{A} + \ri \mathbf{B},
$
then
$\mathbf{p}\cdot\mathbf{p}=0$ means that
\begin{eqnarray}
\mathbf{A}\cdot\mathbf{A} &=& \mathbf{B}\cdot\mathbf{B}, \nn\\
\mathbf{A}\cdot\mathbf{B} &=& 0.
\end{eqnarray}
The first line of the above equation means that vectors $\mathbf{A}$ and $\mathbf{B}$ have the same length; whereas the second line requires that they are perpendicular to each other \cite{footnote4}. We may always rotate the frame such that $\mathbf{A}$ and $\mathbf{B}$ are in the $xy$-plane. That is
$
p_3=0.
$
In this frame, $\mathbf{A}\perp\mathbf{B}$ means that
\begin{equation}
A_1 = \mp B_2,\quad \mathrm{and} \quad A_2 = \pm B_1,
\end{equation}
which is equivalent to
$
p_1 = \pm \ri p_2
$
in terms of the original complex variables. Without loss of generality, let us choose the upper sign, $p_1 = \ri p_2$. (If the lower sign is more convenient, one simply solves $b(t)$ first.)

\subsubsection{$\mathbf{q}\cdot\mathbf{q}\neq0$}

If $\mathbf{q}\cdot\mathbf{q}\neq0$, the solutions are Bessel functions. To see this, let us change variables
\begin{equation}
z(t) \equiv \frac{\ri}{\omega} \sqrt{\mathbf{q}\cdot\mathbf{q}} \eiot , \qquad
a(z) \equiv a(t).
\end{equation}
Then $a(z)$ satisfies a Bessel equation (see \textbf{10.13.1} of Ref.~\cite{DLMF}),
\begin{equation}
a''(z) + \left(1-\frac{2\mathbf{p}\cdot\mathbf{q}}{z^2\omega^2}\right) a(z)=0.
\end{equation}
The solutions are
\begin{equation}
a(t) = C_1 \sqrt{z}J_\nu\left(z\right) + C_2 \sqrt{z} Y_{\nu}\left(z\right),
\end{equation}
with
$
\nu \equiv \sqrt{\frac{2\mathbf{p}\cdot\mathbf{q}}{\omega^2} + \frac{1}{4}}.
$
One can then use this explicit solution to investigate the hopping. One may also analyse this by use of the critical exponent in Eq.~(\ref{gc52}).  We will not repeat the details here as they are much similar to our discussions in previous sections.

\subsubsection{$\mathbf{q}\cdot\mathbf{q}=0$}

If both $\mathbf{p}\cdot\mathbf{p}=0$ and $\mathbf{q}\cdot\mathbf{q}=0$, then the solutions are simple power functions.
\begin{eqnarray}
a(t) &=& C_1 \e^{\frac{\ri \omega t}{2}} + C_2 \exp\left(\frac{\ri \omega t}{2} - \ri t \sqrt{2 \mathbf{q}\cdot\mathbf{q} + \frac{\omega^2}{4}}\right). \nn\\
\end{eqnarray}
Evidently, this special solution has no sudden-switch behavior.

In all these cases, our efforts to explicitly work out the critical exponent help us to find conditions under which the complicated non-Hermitian cycling problem can admit exactly solvable solutions.  Many of these exact solutions might not necessarily further enhance our understanding of piecewise adiabatic following. However, they do clearly indicate that piecewise adiabatic following can occur in multi-frequency driving cases that are even exactly solvable. If the problem is not exactly solvable after all, then we can still resort to the critical exponents emerging from our asymptotic analysis to predict and understand the sudden-switch behaviors.

\section{Conclusion}
In this work, we have extensively investigated the interesting adiabatic following dynamics in periodically driven non-Hermitian systems.  The central concern is the peculiar behavior of cyclic (Floquet) states in the slow-driving limit. It is found that the cyclic states can either behave as intuitively expected by following instantaneous eigenstates of the non-Hermitian Hamiltonian, or exhibit sudden-switching between the instantaneous eigenstates.  As learned from several categories of models under different driving scenarios, the sudden switches from following one instantaneous eigenstate to the other eigenstate can be analyzed or predicted by a universal route -- the sign change of the critical exponent in our asymptotic analysis, which suggests a switch between two terms in the solution with different exponential behavior.  In doing so, we have also discovered many exact solutions in a great variety of non-Hermitian cycling models.  We hope that the many exact solutions found by us can be a useful reference in their own right.   Our next task is to find potential applications of piecewise adiabatic following dynamics in non-Hermitian systems.

\section*{Acknowledgments}
Q.W.~would like to thank Mr.~Jiawen Deng for useful discussion. J.G.~is supported by Singapore Ministry of Education Academic Research Fund Tier I (WBS No.~R-144-000-353-112) and by the Singapore NRF grant No. NRF-NRFI2017-04 (WBS No. R-144-000- 378-281).  Q.W.~is supported by Singapore Ministry of Education Academic Research Fund Tier I (WBS No.~R-144-000-352-112).

\appendix
\section{Time-independent rotation to achieve $p_1-\ri p_2=0$}
\label{sec:appendixA}

In this appendix, we use two methods to prove that $p_1-\ri p_2=0$ can always be achieved by a time-independent rotation.

Algebraically, an arbitrary three-dimensional rotation can be parametrized as
\begin{equation}
R(\alpha,\beta,\gamma) = \e^{\ri\alpha\sigma_3/2} \e^{\ri\beta\sigma_1/2} \e^{\ri\gamma\sigma_3/2}.
\end{equation}
The upper-right corner of the rotated matrix $R^{-1} \mathbf{p}\cdot\bm{\sigma} R$ has the form
\begin{eqnarray}
(p_1-\ri p_2) &\to& \left[(p_1-\ri p_2\cos\beta)\cos\alpha -\ri(p_1\cos\beta-\ri p_2)\sin\alpha\right. \nn\\
&&\quad \left.+ \ri p_3\sin\beta \right]\e^{-\ri\gamma}.
\end{eqnarray}
The complex equation
\begin{equation}
(p_1-\ri p_2\cos\beta)\cos\alpha -\ri(p_1\cos\beta-\ri p_2)\sin\alpha + \ri p_3\sin\beta=0
\end{equation}
can always be solved by proper choices of the two real angle $\alpha$ and $\beta$. Similarly, one can choose a different pair of angles $\alpha$ and $\beta$ such that $(q_1-\ri q_2)=0.$

Alternatively, a new basis with $p_1-\ri p_2=0$ can be found by geometrical means. To see this, let us first write a complex vector as
$
\mathbf{p} = \mathbf{A} + \ri \mathbf{B}.
$
Then the relation  $(p_1-\ri p_2)=0$ leads to
\begin{equation}
A_1 = -B_2, \quad {\rm and} \quad A_2 = B_1,
\end{equation}
which are equivalent to the condition that the vector $(A_1,A_2,0)$ is perpendicular to $(B_1,B_2,0)$.

Here is the procedure to find a new basis with $(A_1,A_2,0)\perp(B_1,B_2,0)$. Consider three planes intersected at the origin, let us label the crossing line between Plane 1 and Plane 2 as Line-12, the crossing line between Plane 1 and Plane 3 as Line-13, and that between Plane 2 and Plane 3 as Line-23. Three crossing lines meet at the origin. First, we request that Planes 1 and 2 be perpendicular to each other.  Then the angle between Line-13 and Line-23 is in general less than $\pi/2$. Second, we adjust Plane 3 such that the angle between Line-13 and Line-23 equals the angle between $\mathbf{A}$ and $\mathbf{B}$ if $\mathbf{A} \cdot \mathbf{B}>0$, or the angle  between $\mathbf{A}$ and $-\mathbf{B}$ if $\mathbf{A} \cdot \mathbf{B}<0$. Finally, we rotate these three planes together to make Line-13 coincide with $\mathbf{A}$ and Line-23 coincide with $\mathbf{B}$ ($-\mathbf{B}$) if $\mathbf{A} \cdot \mathbf{B}>0$ ($\mathbf{A} \cdot \mathbf{B}<0$).  The final step is to define Line-12 as the new $z$-axis, and the plane normal to this $z$ axis
 (not Plane 3) as the $xy$-plane. Then both $\mathbf{A}$ and $\mathbf{A}_\perp=(A_1,A_2,0)$ are in Plane 1, and both $\mathbf{B}$ and $\mathbf{B}_\perp=(B_1,B_2,0)$ are in Plane 2. In this new coordinate system, $(A_1,A_2,0)\perp(B_1,B_2,0)$.  Fig.~\ref{fig:ABz} illustrates our procedure.

\begin{figure}[h!]
	\begin{center}
		\includegraphics[width=0.6\columnwidth]{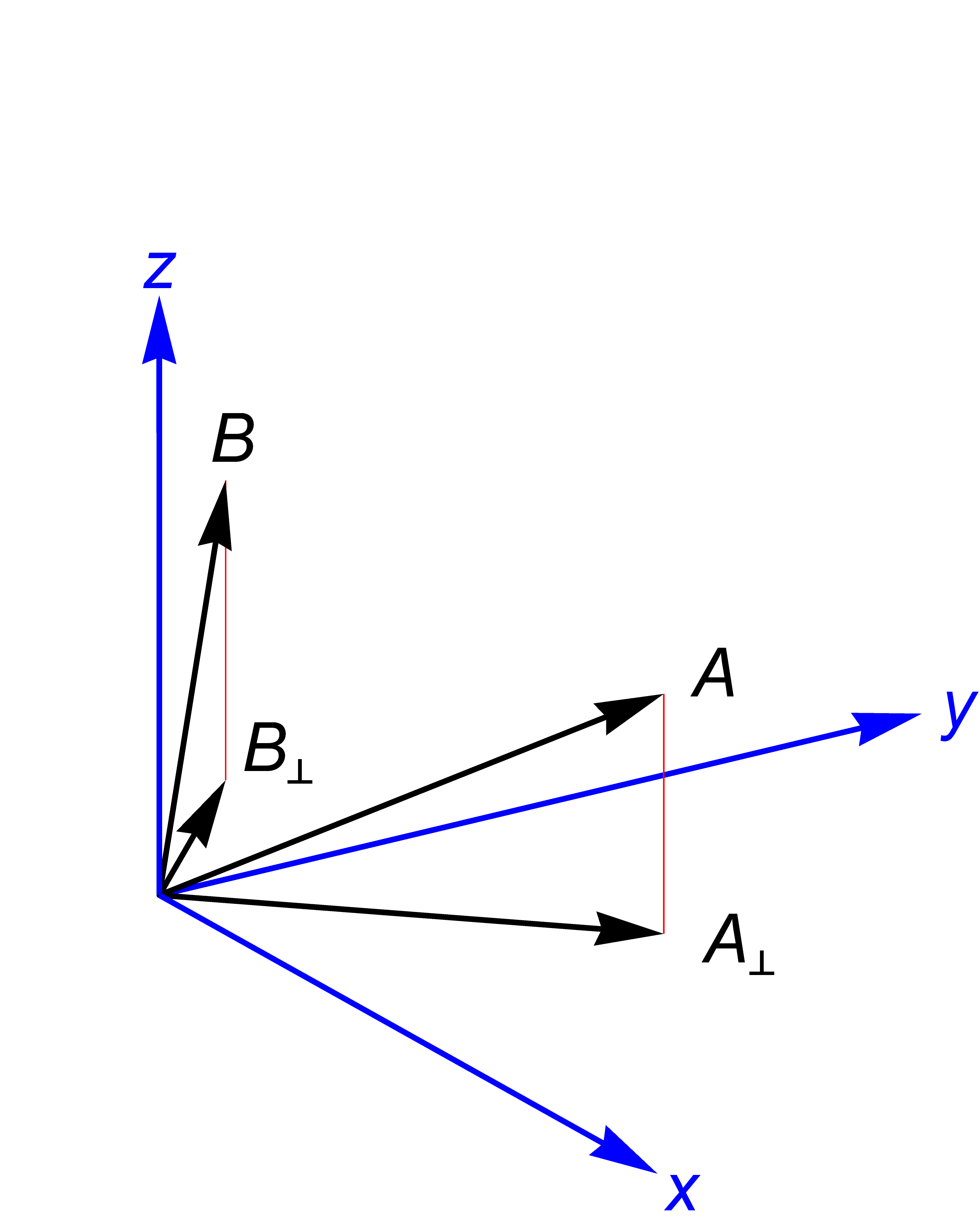}
		\caption{\label{fig:ABz}(color online) Choose a new coordinate system such that $\mathbf{A}_\perp=(A_1,A_2,0)$ is perpendicular to $\mathbf{B}_\perp=(B_1,B_2,0)$.
		}
	\end{center}
\end{figure}

\section{Models with two different frequencies and a constant term}

In this appendix we aim to use the same philosophy to treat the most complicated situations with essentially three frequencies (one of them being zero).  In particular, we may model a traceless model with double nonzero frequencies as
\begin{equation}
H_{0mn}(t) = \mathbf{p}\cdot\bm{\sigma} + \mathbf{q}\cdot\bm{\sigma}\, \e^{\ri m \omega t} + \mathbf{s}\cdot\bm{\sigma}\, \e^{\ri n\omega t},
\end{equation}
where 18 real parameters of the model are organized into three complex vectors, $\mathbf{p}$, $\mathbf{q}$, and $\mathbf{s}$.

\subsection{$n=-m$}

{In this subclass, we first rescale $\omega\to\omega/m$. Then, for convenience, we define
	\begin{equation}
	H_{-101}(t) = \mathbf{s}\cdot\bm{\sigma}\, \e^{-\ri \omega t} + \mathbf{p}\cdot\bm{\sigma} + \mathbf{q}\cdot\bm{\sigma}\, \e^{\ri \omega t}
	\end{equation}
	We change variables as before, $Z=\eiot$ and let $a(Z)=a(t) = \e^{\frac{m(Z)}{\omega}}$. In the long time limit $\omega\to0$, \Eqn{eqn:m(z)} becomes}
\begin{equation}
m(Z)\sim \pm\int^Z\rd x\frac{\sqrt{R(x)}}{x^2},
\label{eqn:m(z)-101}
\end{equation}
where $R(x)$ is quartic instead of quadratic in general,
\begin{equation}
R(x) = \mathbf{s}\cdot\mathbf{s} + 2\mathbf{p}\cdot\mathbf{s}\, x + \left(\mathbf{p}\cdot\mathbf{p} + 2 \mathbf{q}\cdot\mathbf{s}\right) x^2 + 2 \mathbf{p}\cdot\mathbf{q}\, x^3 + \mathbf{q}\cdot\mathbf{q}\, x^4.
\label{eqn:R4}
\end{equation}
This integral can be easily evaluated if
\begin{equation}
\mathbf{s}\cdot\mathbf{s}=0\quad \mathrm{and}\quad \mathbf{p}\cdot\mathbf{s}=0.
\label{eqn:s.s=0}
\end{equation}

Note that an alternative way to obtain a quadratic $R(x)$ is to set $\mathbf{q}\cdot\mathbf{q}=0$ and $\mathbf{p}\cdot\mathbf{q}=0$. This is actually an equivalent set-up if one adopts $\e^{-\ri\omega t}=Z$ as a change of the variables.

\subsubsection{Solvable Double-frequency models with $s_1=\ri s_2$, $s_3=0$, and $\mathbf{p}\cdot\mathbf{s}=0$.}

Having a quadratic $R(x)$ not only makes \Eqn{eqn:m(z)-101} integrable, but also renders the model solvable. The ``master" equation about $a(Z)=a(t)$ is
\begin{eqnarray}
&&\omega^2  Z^2 a''(Z)\nn\\
&=& \left[(p_3 + \omega )p_3 + 2 \mathbf{q}\cdot\mathbf{s} + 2 \mathbf{p}\cdot\mathbf{q}\, Z +  \mathbf{q}\cdot\mathbf{q}\,Z^2\right]a(Z). \nn\\
\label{eqn:Whittaker2}
\end{eqnarray}
The analysis to \Eqn{eqn:Whittaker2} is very similar to that to \Eqn{eqn:Whittaker}. Depending on whether none of, one of, or both of $\mathbf{p}\cdot\mathbf{q}$ and $\mathbf{q}\cdot\mathbf{q}$ vanish, the solutions to \Eqn{eqn:Whittaker2} are Whittaker functions, Bessel functions, or exponential functions, respectively. For example, if none of the two dot products vanishes, the solution of $a(t)$ is Whittaker functions,
\begin{equation}
a(t) = C_1 W_{\kappa,\mu} \left(\frac{2\sqrt{\mathbf{q}\cdot\mathbf{q}}}{\omega}\eiot\right) + C_1 M_{\kappa,\mu} \left(\frac{2\sqrt{\mathbf{q}\cdot\mathbf{q}}}{\omega}\eiot\right),
\end{equation}
with
$
\kappa \equiv -\frac{\mathbf{p}\cdot\mathbf{q}}{\omega\sqrt{\mathbf{q}\cdot\mathbf{q}}}$ and $
\mu \equiv \frac{\sqrt{(2p_3 +\omega)^2 + 8 \mathbf{q}\cdot\mathbf{s}}}{2\omega}.
$
The corresponding $b(t)$ is determined by the same equation in Eq.~(\ref{eqn:b(t)}).

There is a particular interesting example with
\begin{eqnarray}
&& p_1 = \half u_1, \quad p_2 = - \ri \half u_1,\quad p_3= u_3,\nn\\
&& q_1 = \half v_1, \quad q_2 = \ri \half v_1,\quad q_3= v_3,\nn\\
&& s_1 = \half v_2, \quad s_2 = -\ri \half v_2,\quad s_3= 0.
\end{eqnarray}
In the case, the Hamiltonian has the form
\begin{equation}
h_1(t) = \left(
\begin{array}{cc}
u_3+v_3\eiot & v_1\eiot\\
u_1 + v_2 \e^{-\ri \omega t} & -u_3 - v_3\eiot
\end{array}\right).
\end{equation}
This Hamiltonian is equivalent to the single frequency model in Sec.~\ref{sec:SFII} with
\begin{eqnarray}
&& p_1 = \half (v_1+v_2), \quad p_2 = \ri \half (v_1 -v_2),\quad p_3= u_3,\nn\\
&& q_1 = \half u_1, \quad q_2 = -\ri \half u_1,\quad q_3= v_3.
\end{eqnarray}
\begin{equation}
h_2(t) = \left(
\begin{array}{cc}
u_3+v_3\eiot & v_1\\
u_1 \eiot + v_2 & -u_3 - v_3\eiot
\end{array}\right).
\end{equation}
The two Hamiltonians are linked by a simple gauge transformation,
\begin{equation}
h_1(t) = \e^{\frac{\ri\omega t}{2} \sigma_3} h_2(t) \e^{-\frac{\ri\omega t}{2} \sigma_3}
\end{equation}

\subsection{$n=2m$}
{For this subclass, we first rescale $\omega\to\omega/m$. Then, for convenience, we define
	\begin{equation}
	H_{012}(t) = \mathbf{s}\cdot\bm{\sigma} + \mathbf{p}\cdot\bm{\sigma}\, \e^{\ri \omega t} + \mathbf{q}\cdot\bm{\sigma}\, \e^{2\ri \omega t}
	\end{equation}
	We change variables as before, $Z=\eiot$ and let $a(Z)=a(t) = \e^{\frac{m(Z)}{\omega}}$. In the long time limit $\omega\to0$, $ m(Z)$ becomes
	\begin{equation}
	m(Z)\sim \pm\int^Z\rd x\frac{\sqrt{R(x)}}{x},
	\end{equation}
	with $R(x)$ to be identical as in \Eqn{eqn:R4}.}
This integral can also be handled  if
$
\mathbf{q}\cdot\mathbf{q}=0$ and $\mathbf{p}\cdot\mathbf{q}=0,
$
which requires
$
q_1 = \pm \ri q_2$, and $q_3=0.
$
Without loss of generality, let us choose the upper sign, $q_1 = \ri q_2$. (If the lower sign is more convenient, one simply solves $b(t)$ first.) In this frame, $\mathbf{p}\cdot\mathbf{q}=0$ means that
$
p_1=\ri p_2.
$

Note that an alternative way to obtain a quadratic $R(x)$ is to set $\mathbf{s}\cdot\mathbf{s}=0$ and $\mathbf{p}\cdot\mathbf{s}=0$. However, in this case, the model is still hard to solve.

\subsubsection{Solvable Double Frequency Models with $q_1=\ri q_2$, $q_3=0$, and $\mathbf{p}\cdot\mathbf{q}=0$}
Consider one example with {$s_1\neq \ri s_2$ and $p_3^2+\mathbf{q}\cdot\mathbf{s}\neq0$. If $s_1\neq \ri s_2$, then the previous ``master" equation about $a(Z)=a(t)$ with $Z=\eiot$ becomes
\begin{eqnarray}
&& \omega^2  \left[Z^2 a''(Z) + Z a'(Z)\right] \nn\\
&=& \left[\mathbf{s}\cdot\mathbf{s} + (2 \mathbf{p}\cdot\mathbf{s}-\omega p_3)\, Z + (p_3^2+2 \mathbf{q}\cdot\mathbf{s})\,Z^2\right]a(Z). \nn\\
\label{eqn:DF1}
\end{eqnarray}
If $p_3^2+\mathbf{q}\cdot\mathbf{s}\neq0$, the above equation can be transformed into a confluent hypergeometric equation by changing variables
\begin{equation}
z(t) \equiv  \frac{2}{\omega}\sqrt{p_3^2+2\mathbf{q}\cdot\mathbf{s}} \,\eiot, \qquad w(z) \equiv \e^{\half\eiot} \e^{-\ri \sqrt{\mathbf{s}\cdot\mathbf{s}} t} a(t).
\end{equation}
Then $w(z)$ satisfies
\begin{eqnarray}
&& z w''(z) +\left(1+\frac{2}{\omega}\sqrt{\mathbf{s}\cdot\mathbf{s}} - z\right) w'(z)\nn\\
&=& \left[\frac{1}{2} - \frac{p_3}{2 \sqrt{p_3^2+2\mathbf{q}\cdot\mathbf{s}}} + \frac{1}{\omega} \left( \sqrt{\mathbf{s}\cdot\mathbf{s}} + \frac{\mathbf{p}\cdot\mathbf{s}}{\sqrt{p_3^2+2\mathbf{q}\cdot\mathbf{s}}} \right)\right] w(z). \nn\\
\end{eqnarray}
The solutions are the confluent hypergeometric functions
\begin{eqnarray}
a(t) &=& \e^{-\half\eiot} \e^{\ri \sqrt{\mathbf{s}\cdot\mathbf{s}} t}  \left[ C_1 U\left(a,b,\frac{2}{\omega}\sqrt{p_3^2+2\mathbf{q}\cdot\mathbf{s}} \,\eiot\right)\right.\nonumber\\
&&\qquad\left. + C_2\, {}_1F_1\left(a,b,\frac{2}{\omega}\sqrt{p_3^2+2\mathbf{q}\cdot\mathbf{s}} \,\eiot\right) \right],
\end{eqnarray}
with
\begin{eqnarray}
a &\equiv& \frac{1}{2} - \frac{p_3}{2 \sqrt{p_3^2+2\mathbf{q}\cdot\mathbf{s}}} + \frac{1}{\omega} \left( \sqrt{\mathbf{s}\cdot\mathbf{s}} + \frac{\mathbf{p}\cdot\mathbf{s}}{\sqrt{p_3^2+2\mathbf{q}\cdot\mathbf{s}}} \right),\nonumber\\
b &\equiv& 1+\frac{2}{\omega}\sqrt{\mathbf{s}\cdot\mathbf{s}}.
\end{eqnarray}

As a second example, consider $p_3^2+\mathbf{q}\cdot\mathbf{s}=0$ and $s_1\neq \ri s_2$. Then if $p_3^2+\mathbf{q}\cdot\mathbf{s}=0$, but $s_1\neq \ri s_2$, the ``master" equation becomes a Bessel equation again. The solutions of $a(t)$ is
\begin{eqnarray}
a(t) &=& C_1 J_\nu\left(\frac{2\ri\sqrt{2\mathbf{p}\cdot\mathbf{s}-\omega p_3}}{\omega} \e^{\frac{\ri \omega t}{2}}\right) \nn\\
&&\quad + C_2 J_{-\nu}\left(\frac{2\ri\sqrt{2\mathbf{p}\cdot\mathbf{s}-\omega p_3}}{\omega} \e^{\frac{\ri \omega t}{2}}\right)
\end{eqnarray}
with $\nu \equiv \frac{2\sqrt{\mathbf{s}\cdot\mathbf{s}}}{\omega}.$
	
As a final solvable example, consider {$s_1=\ri s_2$}. Then the differential equations reduce to the first-order ones. The solution of $a(t)$ is exponential function and that of $b(t)$ is related to the incomplete Gamma function.



\begin{thebibliography}{90}
\bibitem{footnote1}
Note that such a sudden change will never occur in a unitary evolution. The effective Hamiltonian therein accounting for the time evolution from the start to arbitrary later times is always Hermitian. Thus the critical exponents there are always purely imaginary. The Stokes phenomenon requires a switch in the signs of the real part of the critical exponent. Therefore there is no Stokes phenomenon in a unitary evolution.
\bibitem{footnote4}
This means that the matrix $\mathbf{p}\cdot\bm{\sigma}$ is $\mathcal{PT}$-symmetric \cite{WCZ}.
\end{thebibliography}
\end{document}